\documentclass[12pt,twoside]{article}

\pdfoutput=1

\usepackage[usenames]{color}
\usepackage{lscape}
\usepackage{amssymb}
\usepackage{rotating}
\usepackage[T1]{fontenc}
\usepackage{graphicx}
\usepackage{fancyhdr}
\usepackage{amsmath}
\usepackage{ae}
\usepackage{aecompl}
\usepackage{hyperref}

\def\nn{\nonumber \\}

\newlength{\dinwidth}
\newlength{\dinmargin}
\begin{document}

\thispagestyle{empty}
\begin{flushright}
Cavendish--HEP--11/16\\
DAMTP--2011--52\\
IFT--11--07\\
\end{flushright}

\vspace*{1cm}

\centerline{\Large\bf Yukawa unification in SO(10)} 
\vspace*{2mm}
\centerline{\Large\bf with light sparticle spectrum}

\vspace*{5mm}

\vspace*{5mm} \noindent
\vskip 0.5cm
\centerline{\bf
Marcin Badziak${}^{a,b,}$\footnote[1]{M.Badziak@damtp.cam.ac.uk},
Marek Olechowski${}^{c,}$\footnote[2]{Marek.Olechowski@fuw.edu.pl},
Stefan Pokorski${}^{c,}$\footnote[3]{Stefan.Pokorski@fuw.edu.pl}
}
\vskip 5mm
\begin{center}
{${}^a$\em Department of Applied Mathematics and Theoretical Physics, 
Centre for Mathematical Sciences, University of Cambridge, 
Wilberforce Road, Cambridge CB3 0WA, United Kingdom \\[5mm]

${}^b$Cavendish Laboratory, University of Cambridge, J.J. Thomson Avenue, \\ Cambridge CB3 0HE, United Kingdom}
\end{center}
\centerline{${}^c$\em Institute of Theoretical Physics,
University of Warsaw} 
\centerline{\em ul.\ Ho\.za 69, PL--00--681 Warsaw, Poland}

\vskip 1cm

\centerline{\bf Abstract}
\vskip 3mm

We investigate supersymmetric SO(10) GUT model with $\mu<0$. 
The requirements of top-bottom-tau Yukawa unification, correct 
radiative electroweak symmetry breaking and agreement with the 
present experimental data may be met when the soft masses of 
scalars and gauginos are non-universal. We show how appropriate 
non-universalities can easily be obtained in the SO(10) GUT broken 
to the Standard Model. We discuss how values of BR$(b\to s\gamma)$ 
and $(g-2)_\mu$ simultaneously in a good agreement with the 
experimental data can be achieved in SO(10) model with $\mu<0$. 
In the region of the parameter space preferred by our analysis 
there are two main mechanisms leading to the LSP relic abundance 
consistent with the WMAP results. One is the co-annihilation with 
the stau and the second is the resonant annihilation via exchange 
of the $Z$ boson or the light Higgs scalar. 
A very interesting feature of SO(10) models with 
negative $\mu$ is that they predict relatively 
light sparticle spectra. 
Even the heaviest superpartners may easily have masses 
below 1.5 TeV in contrast to multi-TeV particles typical  
for models with positive $\mu$.

\newpage

\section{Introduction}

Supersymmetric Grand Unified Theory (SUSY GUT) based on the SO(10) 
gauge group is arguably one of the most attractive theories which 
go beyond the Standard Model (SM). In the simplest version of 
SO(10) SUSY GUT, the Minimal Supersymmetric Standard Model 
(MSSM) Higgs doublets, $H_u$ and $H_d$, sit in one 10-dimensional 
representation, while all fermions (together with the right-handed 
neutrino) of each generation belong to one 16-dimensional 
representation. In such model not only the gauge couplings but also 
the Yukawa couplings of top, bottom and tau unify at the GUT scale.
This condition together with the requirement of radiative electroweak 
symmetry breaking (REWSB) strongly restricts the allowed pattern of 
soft supersymmetry breaking terms. In particular, top-bottom-tau
Yukawa unification cannot be obtained if the universal boundary 
conditions for the soft terms are assumed at the GUT scale \cite{Carena}.

One possibility that makes the top-bottom-tau Yukawa unification 
viable is some splitting between the soft masses of the Higgs 
doublets and/or among the third generation squarks \cite{Olechowski}. 
Since $H_u$ and $H_d$ are parts 
of the same SO(10) multiplet, Higgs masses cannot be split by hand. 
The same applies to sfermions of each generation. 
However, splitting within SO(10) multiplets can be consistent 
with SO(10) gauge symmetry if $D$-term contributions to the scalar 
masses are taken into account. Such contributions are generically 
present in effective theories resulting from SO(10) GUT models 
\cite{Dterm} and they simultaneously split soft masses of 
Higgses, squarks and sleptons.

Requirement of top-bottom-tau Yukawa unification prefers negative 
values of the Higgs-mixing parameter,
$\mu$, (negative relative to the SU(3) gaugino mass, $M_3$) 
because in such a case the MSSM threshold correction to the bottom quark 
mass has generically sign appropriate for bottom-tau Yukawa unification. 
Indeed, it was shown in  \cite{Baernegmu}  that for $\mu<0$ top-bottom-tau 
Yukawa unification can be obtained with universal gaugino masses and 
$D$-term splitting of scalar masses. However, that model predicts a 
negative SUSY contribution to the muon anomalous magnetic moment, 
$(g-2)_{\mu}$, and enlarges the observed discrepancy between theory 
and experiment. This problem can be solved if the SU(2) gaugino mass, 
$M_2$, is negative.

In this article, we investigate models with negative $\mu$ and 
non-universal soft masses of scalars and gauginos consistent 
with SO(10) symmetry.  
First, we split the soft scalar masses at the GUT scale using
$D$-terms appearing when SO(10) is broken to the SM gauge 
symmetry group. We do not introduce any intergenerational 
splitting at the GUT scale. Second, we assume that the gaugino 
masses are generated by an $F$-term which is a non-singlet of SO(10)  
transforming as 24-dimensional representation of SU(5) $\supset$ SO(10). 
This assumption results in the following pattern of the gaugino 
masses: $M_1:M_2:M_3=-1:-3:2$ \cite{Martin}, which implies that 
the SUSY contribution to $(g-2)_{\mu}$ is positive, as preferred by the  
experimental data.
We should stress here that even though the gaugino masses are non-universal, 
there is only one free parameter in this sector which sets the overall 
scale.

We show that in the above framework top-bottom-tau Yukawa unification 
can be achieved. We identify some correlations between the GUT scale 
values of the soft supersymmetry breaking parameters which are required 
to keep the SUSY threshold correction to the bottom mass small enough 
to be consistent with Yukawa unification. Given that these correlations 
hold, Yukawa-unified solutions exist for a very wide range of the gaugino 
and scalar masses. The main constraints on this model follow from the 
experimental bounds on BR$(b \to s \gamma)$, $(g-2)_{\mu}$ and the 
dark matter relic abundance. 
We find that the combination of these constraints, together with 
the condition of top-bottom-tau Yukawa unification, leads to a rather 
definite prediction for the SUSY spectrum. In particular, given that  
BR$(b \to s \gamma)$ and $(g-2)_{\mu}$ are compatible with the experimental 
data at the $2\sigma$ level, the gluino mass is predicted to be between 
$500$ and $700$ GeV or between $900$ GeV and $1.6$ TeV, while all other 
sparticles have masses below $2$ TeV. This is the first SO(10) model 
in which top-bottom-tau Yukawa unification is consistent with such 
relatively light SUSY spectrum.

Some aspects of top-bottom-tau Yukawa unification in 
supersymmetric models has been investigated before. 
It has been shown \cite{Blazek,Auto,BaerDM} that the 
precise unification cannot be obtained with 
$D$-term splitting of scalar masses if $\mu$ is positive. 
The level of Yukawa unification improves somewhat 
if at the GUT scale the third generation scalar masses are different 
than the scalar masses of the first and the second generation and/or 
the effect of right-handed neutrinos on the renormalization group equations 
is taken into account. However, even in such a case the remaining 
discrepancy \cite{Baerjustso} may be hard to explain by GUT scale 
threshold corrections to the Yukawa couplings. On the other hand, 
Yukawa-unified solutions for $\mu>0$ were found if only Higgs masses 
were split and soft masses of squarks and sleptons remained universal 
at GUT scale \cite{Blazek,Auto,BaerDM}. Beside the fact that this kind 
of splitting explicitly breaks the SO(10) symmetry, 
that scenario suffers also from phenomenological drawbacks. 
Namely, it predicts multi-TeV scalar masses and thermal relic abundance 
of neutralinos several orders of magnitude larger than the upper 
experimental bound.

Models with $\mu<0$ may be consistent with the constraints on  
BR$(b\to s\gamma)$ and $(g-2)_{\mu}$ only if $M_2$ is negative 
\cite{ChNath}. This observation\footnote{
Top-bottom-tau Yukawa unification in a model with $\mu<0$ and non-universal
gaugino masses assumed to be generated by an $F$-term in 54-dimensional
representation of SO(10) was investigated before in \cite{ChCoNath}. 
However, the gaugino mass relation used in that analysis was based 
on the results of \cite{wronggaugino} which are incorrect, as was 
pointed out recently in \cite{Martin}. Moreover, the Higgs splitting 
in the model of \cite{ChCoNath} was introduced ad-hoc rather than by 
a $D$-term contribution.} 
was recently used in \cite{GoKhRaSh} where top-bottom-tau Yukawa 
unification was considered in the context of supersymmetric 
$SU(4)_c\times SU(2)_L\times SU(2)_R$. It was found that assuming 
$\mu<0$ and $M_2<0$ and varying $M_2$ and $M_3$ independently, 
top-bottom-tau Yukawa unification can be obtained without a conflict 
with any experimental data. However, in \cite{GoKhRaSh} an ad-hoc Higgs 
mass splitting was used which in the case of SO(10), considered in this 
paper, has no theoretical justification.

The paper is organized as follows: The importance of the sign of $\mu$ 
on the Yukawa unification is discussed in Section \ref{sec_mu}. 
Possible patterns of non-universalities of the soft terms  
are presented in Section \ref{sec_soft_terms}. Section \ref{sec_bsg_g2} 
contains a detailed discussion of the MSSM contributions to
BR$(b\to s\gamma)$ and $(g-2)_\mu$. The impact of the experimental 
bounds on these two quantities on the SUSY parameter space is 
also analyzed. Some important facts concerning the relic abundance 
of the neutralinos are recalled in Section \ref{sec_LSP}. 
Section \ref{sec_numerical_results} is devoted to the results 
obtained by our numerical analysis. Finally, we conclude in Section 
\ref{sec_conclusions}.

\section{\texorpdfstring{SO(10) Yukawa unification with negative $\boldsymbol{\mu}$}{SO(10) Yukawa unification with negative mu}}
\label{sec_mu}

Due to the large mass of the top quark, the top-bottom Yukawa unification 
requires a large value of $\tan\beta\sim m_t/m_b$. This in turn has a 
crucial impact on the bottom-tau Yukawa unification because 
for large $\tan\beta$ there are sizable supersymmetric 
corrections to the bottom quark mass \cite{Carena,Hall,Pierce}:
\begin{equation}
m_b^{\rm MSSM}
=m_b^{\rm SM}\left[1+\left(\frac{\delta m_b}{m_b}\right)\right]^{-1}
.
\end{equation}
Unification of the bottom and tau Yukawa couplings requires a negative 
correction to the bottom mass. The logarithmic correction 
is positive and of order ${\mathcal O}(5\%)$.
Therefore, the finite correction has to be necessarily negative and 
large enough to compensate the logarithmic one. The main correction  
originates from gluino-sbottom and chargino-stop loops 
\cite{Carena,Hall,Pierce}: 
\begin{equation}
\label{mbsusycorr}
\left(\frac{\delta m_b}{m_b}\right)^{\rm finite}
\approx\frac{g_3^2}{6\pi^2}\mu m_{\tilde g}\tan\beta \, 
I(m_{\tilde b_1}^2,m_{\tilde b_2}^2,m_{\tilde g}^2)
+\frac{h_t^2}{16\pi^2}\mu A_t\tan\beta \, 
I(m_{\tilde t_1}^2,m_{\tilde t_2}^2,\mu^2) \,,
\end{equation}
where the loop integral $I(x,y,z)$ is defined e.g.\ in the appendix 
of \cite{Hall}. What is relevant for us is that this integral scales 
as the inverse of the mass squared of the heaviest particle propagating 
in the loop and it can be approximated by 
$I(x,y,z)\approx a/\max(x,y,z)$ with $a$ typically between $0.5$ and $1$. 
The numerical coefficient in front of the gluino-sbottom contribution 
is significantly larger than the one in front of the chargino-stop 
contribution, so the former contribution dominates in the vast
majority of the parameter space. 
The gluino-sbottom contribution to the bottom mass (\ref{mbsusycorr})
has the same sign as the sign of the $\mu$ parameter\footnote
{
We use the sign convention in which the gluino mass parameter $M_3$ 
and the gluino mass $m_{\tilde{g}}$ are positive.
}. 
Therefore, Yukawa unification prefers the negative sign of $\mu$.

Since the gluino-sbottom correction to the bottom mass is the 
dominant one let us discuss it in some more detail. It is convenient 
to rewrite it using the experimental value of the strong coupling 
constant, the properties of the loop integral $I(x,y,z)$ appearing 
in (\ref{mbsusycorr}) and the value of $\tan\beta\approx50$ 
(as required by top-bottom unification):
\begin{equation}
\label{mb_gluinocorr}
\left(\frac{\delta m_b}{m_b}\right)^{\tilde{g}\tilde{b}}
\approx {\mathcal O}(0.5 - 1)\frac{\mu}{m_{\tilde g}}\, 
{\rm min}\left(1,\left(\frac{m_{\tilde g}}{m_{\tilde{b}}}\right)^2\right)
,
\end{equation}
where $m_{\tilde{b}}$ is the mass of the heavier sbottom. The magnitude 
of the finite threshold corrections required for bottom-tau unification 
must be about 10\% to 20\% \cite{Wells_yuk}. From eq.\ (\ref{mb_gluinocorr}) 
it is clear that the correction to the bottom mass is far too large  
unless $|\mu|< m_{\tilde g}$ or $m_{\tilde g}<m_{\tilde{b}}$. 
In the case of $m_{\tilde g}\geqslant m_{\tilde{b}}$ one easily finds 
the following (conservative) upper bound for $|\mu|$:
\begin{equation}
 \label{mubound}
|\mu|\lesssim 0.4 m_{\tilde g}\approx M_3 \,,
\end{equation}
where $M_3$ is the gluino mass at the GUT scale. This bound is relaxed 
for $m_{\tilde g}<m_{\tilde{b}}$ 
and e.g. $|\mu|$ can be larger than $m_{\tilde g}$ for the gluino 
mass about two times smaller than the heavier sbottom mass.

Yukawa coupling unification may be achieved also for positive $\mu$. 
However, in such a case, the parameters must satisfy some quite 
strong constraints. First: the gluino-sbottom 
contribution in (\ref{mbsusycorr}) has the ``wrong'' sign, 
so it should be suppressed. This usually requires very large 
soft sfermion masses $m_{16}\gg M_{1/2}$. Second: the soft trilinear 
parameter $A_t$ must be negative and large enough to produce a 
negative correction to the bottom mass compensating 
the positive (logarithmic and gluino-sbottom) contributions. 
This requires the GUT scale parameter $A_0$ to be negative, large 
and carefully chosen (typically $A_0\approx-{\cal O}(2.5)m_{16}$).
This characteristic pattern of large soft parameters required 
by Yukawa unification with positive $\mu$ was found and discussed 
in \cite{Blazek,Auto}.

Of course, there are some constraints on the soft parameters 
also in the case of negative $\mu$. However, they are weaker than in 
models with positive $\mu$. For example: for $\mu>0$, the parameter 
$A_t$ must be large and negative while for $\mu<0$ it can be of both 
signs and with not too large absolute value. The sign and the magnitude 
of the gluino-sbottom contribution in (\ref{mbsusycorr}) is typically 
correct for $\mu<0$ and this should not be spoiled by a too large 
chargino-stop contribution (neither positive nor negative). 
The soft sfermion mass parameter $m_{16}$ must be very large  for 
$\mu>0$ while there is no such requirement in the case of $\mu<0$.
These differences in the constraints have important phenomenological 
implications. The SUSY spectrum is very heavy in models with 
positive $\mu$ but, as we show later, may be relatively light 
for negative $\mu$.

The sign of $\mu$ is important also for other features of SO(10) models. 
For example the sign of the dominant SUSY contribution to the muon 
anomalous magnetic moment, $(g-2)_\mu$, is the same as the sign of 
the product $\mu M_2$. This contribution should be positive because 
the SM prediction for $(g-2)_\mu$ is significantly smaller than the 
experimentally measured value. Thus, extentions of the SM with a negative 
contribution to $(g-2)_\mu$ are very strongly disfavored. 
This is the case e.g.\ for SUSY models with negative 
$\mu$ and universal gaugino masses.  
So, in phenomenologically acceptable models with $\mu<0$ one needs  
gaugino masses with negative $M_2$. It will be shown in the next 
section that this situation can be realized in SO(10) GUT models.

It is well-known that for universal soft SUSY breaking terms at 
the GUT scale top-bottom-tau Yukawa unification is incompatible with 
the radiative electroweak symmetry breaking \cite{Carena}. 
The main reason is that at the electroweak scale the soft 
Higgs masses must satisfy the condition
\begin{equation}
m_{H_d}^2 - m_{H_u}^2 > M_Z^2
\,.
\label{mH1mH2}
\end{equation}
For large $\tan\beta$ and heavy 
top quark, the value of $M_{1/2}^2$ is strongly correlated with that 
of $\mu^2$. This correlation implies $\mu^2>M_{1/2}^2$ which is 
inconsistent with the upper bound (\ref{mubound}) on $|\mu|$ and leads 
to a too large (positive or negative) correction to the bottom mass.
As a result, the correct REWSB can not be achieved in SO(10) 
models with universal soft terms at the GUT scale.

The situation changes when non-universalities are allowed. 
Patterns of non-universal scalar masses which make top-bottom-tau 
Yukawa unification compatible with the REWSB were identified in 
\cite{Olechowski}. Two examples of such patterns at the GUT scale are: 
$m_{H_d}^2>m_{H_u}^2>m_0^2$ and $m_{D}^2<m_{U}^2<m_0^2$, where $m_0$ 
is the common value of the soft mass of all particles other than 
$H_d$ and $H_u$ ($D$ and $U$) in the first (second) case. 
As we will show below, similar non-universalities may very 
naturally emerge in the SO(10) GUT models.

\section{Patterns of soft terms allowed in SO(10) models}
\label{sec_soft_terms}

SO(10) GUT is a very predictive theory since its structure allows 
only few free parameters. Both Higgs doublets reside in the 
same representation so they have a common soft SUSY breaking mass, 
$m_{10}$. The same happens with all squarks and sleptons of one 
generation and we denote their mass as 
$m_{16}$ which, in general, can be different from $m_{10}$ 
(there can be three different soft sfermion masses, one for 
each generation, but we do not consider intergenerational 
splitting in this paper).
In addition, in the effective MSSM (below the GUT scale) there is 
another source of scalar masses which may differentiate masses 
within each SO(10) representation. It originates from 
breaking of U(1) which  occurs when 
SO(10) is broken down to SM gauge group. Even though the exact 
mechanism of GUT symmetry breaking is unknown, it is plausible that 
it occurs due to the existence of some SM singlets which are charged 
under the additional U(1) (which is part of SO(10)) and acquire 
vacuum expectation values. Since masses of these SM singlets are 
expected to be around the GUT scale, they can be integrated out. 
It can be shown that in the effective theory the MSSM scalars acquire 
squared mass corrections proportional to their charges under the 
broken U(1) \cite{Dterm}. The magnitude of these corrections is set 
by the $D$-term of the additional U(1). Taking this into account 
one finds the following generic structure of scalar masses in SO(10) 
models:
\begin{align}
\label{scalarDterm}
&m_{H_d}^2=m_{10}^2+2D\,, \nn[4pt]
&m_{H_u}^2=m_{10}^2-2D\,, \nn[4pt]
&m_{Q,U,E}^2=m_{16}^2+D\,, \nn[4pt]
&m_{D,L}^2=m_{16}^2-3D\,, 
\end{align} 
where  $D$ parameterizes the size of a $D$-term contribution.
It was demonstrated in \cite{Murayama} that the above source of
non-universal scalar masses, for a certain range of the parameters 
$m_{16}$ and $m_{10}$, can lead to top-bottom Yukawa unification 
in agreement with REWSB if $D$ is positive\footnote{
For positive $D$, one gets 
$m_{H_d}^2>m_{H_u}^2$ and $m_{D}^2<m_{U}^2$ as in two patterns of 
non-universalities proposed in \cite{Olechowski} 
(mentioned after eq.\ (\ref{mH1mH2})). In  \cite{Murayama}, the parameters 
$m_{16}$ and $m_{10}$ were taken as universal at the Planck scale and 
their values at the GUT scale were obtained by RG running in SO(10).
In the present paper, we take them as free parameters.
}. 
However, such non-universalities 
are not enough if one wants to construct models satisfying all present 
experimental constraints.

In GUT models the gaugino masses are usually assumed to be universal. 
However, they are universal only in the  special case when the SUSY 
breaking $F$-term which gets a vacuum expectation value is a singlet 
of the GUT gauge symmetry group. In general, the gaugino masses 
in supergravity can arise from the following dimension five 
operator:
\begin{equation}
{\cal L}\supset -\frac{F^{ab}}{2 M_{\rm Planck}}
\lambda^a \lambda^b + {\rm c.c.}\,,
\end{equation}
where $\lambda^a$ are the gaugino fields. The resulting gaugino mass 
matrix is $\frac{\langle F^{ab}\rangle}{M_{\rm Planck}}$. The vacuum 
expectation value of the relevant $F$-term, $\langle F^{ab}\rangle$, 
must transform as the singlet of the SM gauge group but  
it can be a non-singlet of the full GUT group. Since the gauginos 
belong to the adjoint representation, non-zero gaugino masses may 
arise from VEVs of the $F$-terms transforming as any of the 
representations present in the symmetric part of the direct 
product of the two adjoints, which for SO(10) reads: 
\begin{equation}
({\bf 45} \times {\bf 45} )_{\rm S} 
= {\bf 1} + {\bf 54} + {\bf 210} + {\bf 770} \,.
\label{45x45}
\end{equation} 
If SUSY is broken by an $F$-term transforming as a non-singlet 
representation of SO(10), gaugino masses are not universal. 
The classification of non-universal gaugino
masses for SO(10) and its subgroups was provided in \cite{Martin}.

From our point of view the most interesting case is when SUSY is 
broken by an $F$-term transforming as the 24-dimensional representation 
of SU(5) $\supset$ SO(10). Such $F$-term gives a negative 
contribution to the gauginos associated with the SU(2) subgroup 
of the SM gauge group: 
\begin{equation}
\label{gauginoratio}
 M_1:M_2:M_3 = -\frac{1}{2}:-\frac{3}{2}:1\,,
\end{equation}
where $M_1$, $M_2$ and $M_3$ are the bino, wino and gluino masses, 
respectively. The overall scale of gaugino masses, $M_{1/2}=M_3$, 
is treated as a free parameter. 

The 24-dimensional representation of $SU(5)$ appears in each of 
the three non-singlet representations of $SO(10)$ in the r.h.s.\ 
of eq.\ (\ref{45x45}). However, the most economical choice is 
the {\bf 54} representation. It is the smallest one and it contains 
only one SM singlet (in {\bf 24} of SU(5)). Representations {\bf 210} 
and {\bf 770} contain 3 and 4 SM singlets, respectively.

In this paper, we assume that the gaugino masses arise predominantly 
from the $F$-term in ${\bf 54}$ representation 
(or a part of ${\bf 210}$ or ${\bf 770}$ 
representation transforming as ${\bf 24}$ of SU(5)) 
so they satisfy the relation (\ref{gauginoratio}).

For simplicity, we assume that soft scalar masses are given by 
(\ref{scalarDterm}) and the soft trilinear terms have 
a universal value $A_0$. The first assumption is realized when 
the non-singlet $F$-term appears in the soft scalar terms
only in the singlet combination (e.g.\ the singlet in the 
product ${\bf 54} \times {\bf 54}$). The latter assumption 
requires the existence of a singlet F-term, in addition to 
a non-singlet one. Such singlet must dominate the trilinear 
terms and may contribute to the scalar masses but should give 
only a subdominant contribution to the gaugino masses. 
The consequences of dropping these two simplifying assumptions 
will be discussed elsewhere.

\section{\texorpdfstring{Interplay between BR$\boldsymbol{(b\to s \gamma)}$ 
and $\boldsymbol{(g-2)_{\mu}}$}{Interplay between BR(b-> s gamma)
and (g-2)}}
\label{sec_bsg_g2}

The experimental bounds on two quantities: 
BR$(b\to s \gamma)$ and $(g-2)_{\mu}$, give quite strong constraints 
on the SUSY extensions of the SM. The reason is as follows: 
The SM prediction \cite{Davier_g2SM} for the anomalous muon magnetic 
moment is more than $3\sigma$ below the present experimental result 
\cite{BNL}:
\begin{eqnarray}
a_\mu^{\rm SM}
&&\!\!\!\!\!\!\!\!
=\left(11659180.2\pm4.9\right)\times 10^{-10} 
\,,
\nn[4pt] 
a_\mu^{\rm exp}
&&\!\!\!\!\!\!\!\!
=\left(11659208.9\pm6.3\right)\times 10^{-10} 
\,,
\end{eqnarray}
where $a_{\mu}\equiv(g-2)_\mu/2$. Combining in quadrature theoretical 
and experimental errors yields the following result for the discrepancy 
between experiment and the SM prediction:
\begin{equation}
 \delta a_{\mu}\equiv a_\mu^{\rm exp}-a_\mu^{\rm SM}
=\left(28.7\pm8.0\right)\times 10^{-10}
\,.
\end{equation}
Therefore, the SUSY contribution to $(g-2)_{\mu}$ should be rather 
big and necessarily positive. On the other hand, the SM 
prediction \cite{Misiak} for BR$(b\to s \gamma)$ is quite close 
to the experimental result \cite{HFAG}:
\begin{eqnarray}
{\rm BR}^{\rm SM}(b\to s \gamma)
&&\!\!\!\!\!\!\!\!
= \left(3.15\pm0.23\right)\times 10^{-4} 
\,,
\nn[4pt] 
{\rm BR}^{\rm exp}(b\to s \gamma)
&&\!\!\!\!\!\!\!\!
=\left(3.55\pm0.24\pm0.09\right)\times
10^{-4}
\,,
\end{eqnarray}
leaving no much room for the SUSY contribution. Moreover, 
the charged Higgs exchange always increases the SM result 
so the contributions involving superpartner loops must 
be small or negative.

Typically, SUSY effects decrease with the SUSY mass scale. 
So, the measured value of $(g-2)_{\mu}$ strongly prefers small 
$M_{\rm SUSY}$ while constraints on BR$(b\to s \gamma)$ can be  
easier fulfilled for bigger $M_{\rm SUSY}$. One should remember that 
the LEP bound on the Higgs mass also requires heavy superpartners 
(at least the stops). This, apparently, leads to a tension between 
different experimental results.

It should be stressed that one should be very careful when using 
$(g-2)_{\mu}$ and BR$(b\to s \gamma)$ to exclude various MSSM scenarios. 
On the one hand, there are still unresolved issues concerning the 
calculation of $(g-2)_{\mu}$ in the Standard Model (such as the 
value of the hadronic contribution) and it may eventually turn out that 
the observed discrepancy between the SM and the experiment is not so big. 
On the other hand, the calculation of BR$(b\to s \gamma)$ in the MSSM 
may be affected by some non-minimal flavour violating effects such as 
non-vanishing off-diagonal elements of soft mass squared matrix for 
the up-type squarks (see e.g.\ \cite{Chankowski},\cite{Okumura}).
Keeping this in mind, we will first consider separately two cases in 
which only one of the two above mentioned constraints is imposed on 
our SO(10) model. Then, we will focus on the question if both 
constraints can be satisfied simultaneously. In the following 
subsections we discuss the main MSSM contribution to $(g-2)_{\mu}$ 
and BR$(b\to s \gamma)$ in some more detail.

\subsection{\texorpdfstring{MSSM contributions to $\boldsymbol{(g-2)_{\mu}}$}{MSSM contributions to (g-2)}}

Two dominant contributions to $(g-2)_{\mu}$ in the MSSM originate from 
the one-loop diagrams involving charginos accompanied by the muon sneutrino 
and neutralinos accompanied by smuons. 
The $\tan\beta$ enhanced part of the chargino-sneutrino contribution 
is given by
\cite{Moroi},\cite{Stockinger}:
\begin{equation}
a_\mu^{\chi^\pm}
\approx
\frac{g_2h_\mu m_\mu}{24\pi^2}
\sum_a \frac{m_{\chi_a^+}}{m_{\tilde{\nu}_\mu}^2}
U_{a2} V_{a1}F_2^C\left(\frac{m_{\chi_a^+}^2}{m_{\tilde{\nu}_\mu}^2}\right)
,
\label{amu-chargino}
\end{equation}
where the loop function $F_2^C$ may be found in \cite{Moroi},
\cite{Stockinger}\footnote{
A factor of 1/2 is missing in the definition of $F_2^C$ 
in ref.\ \cite{Stockinger}.
}. Assuming that all SUSY particles are degenerate with a common 
mass $M_{\rm SUSY}$ and neglecting terms proportional to 
$M_W/M_{\rm SUSY}$, the chargino contribution may be further 
approximated as
\begin{equation}
a_\mu^{\chi^\pm}
\approx
\frac{1}{32\pi^2}\frac{m_\mu^2}{M_{\rm SUSY}^2}
g_2^2\,{\rm sgn}(\mu M_2)\tan\beta 
.
\label{amu-chargino_approx}
\end{equation}
In the same approximation the neutralino-smuon contribution reads
\begin{equation}
a_\mu^{\chi^0}
\approx\frac{1}{192\pi^2}
\frac{m_\mu^2}{M_{\rm SUSY}^2}
\left[g_1^2\,{\rm sgn}(\mu M_1)-g_2^2\,{\rm sgn}(\mu M_2)\right]
\tan\beta 
.
\label{amu-neutralino_approx}
\end{equation}
The numerical coefficient in front of the chargino contribution 
(\ref{amu-chargino_approx}) is several times bigger than the one 
in front of the neutralino contribution (\ref{amu-neutralino_approx}).
Thus, typically the chargino loop dominates the SUSY contribution 
to $a_\mu$. Its value (\ref{amu-chargino}) decreases with 
$m_{\tilde{\nu}_\mu}^2$, so experimental data on $(g-2)_{\mu}$ 
favor light muon sneutrino.

Formulae (\ref{amu-chargino_approx}) and (\ref{amu-neutralino_approx})
are valid in the limit in which all SUSY particles have a common 
mass $M_{\rm SUSY}$. Actually different diagrams involve different 
masses. Thus, in principle the neutralino contribution might be  
dominant if smuons were much lighter than the muon sneutrino. 
However, this is not the case in our model because the $D$-term 
splitting of the scalar masses implies that $\tilde{\mu}_R$ is always 
heavier than the muon sneutrino ($\tilde{\mu}_L$ is almost degenerate 
with muon sneutrino). In our numerical calculations we use full 
expressions for $(g-2)_\mu$ but to discuss qualitatively the results 
it is enough to use only the chargino-sneutrino contribution 
(\ref{amu-chargino}).

\subsection{\texorpdfstring{MSSM contributions to BR$\boldsymbol{(b\to s \gamma)}$}{MSSM contributions to BR(b-> s gamma)}}
\label{sec_bsg}

Within MSSM there are two possibly sizable 
contributions to BR$(b\to s \gamma)$. The first one, involving the loop 
with the charged Higgs boson, always adds constructively to the SM result. 
The second important contribution comes from the diagrams with 
charginos and accompanying squarks in the loop. 
The chargino contribution is enhanced by $\tan\beta$ so it may be 
especially large in models with top-bottom-tau Yukawa unification
considered in this paper.

For large $\tan\beta$ the chargino-squark contribution is dominated 
by the following part of the relevant Wilson coefficients \cite{DeGaGi}
\begin{eqnarray}
C^{\chi^+}_{7,8}
\approx
\frac{1}{\cos\beta}\sum_{a=1,2}
&&\!\!\!\!\!\!
\left\{
\frac{{U}_{a2}{V}_{a1}M_W}{\sqrt{2}m_{\chi_a^+}}
\left[
F^{(3)}_{7,8}\left(\frac{m_{\tilde{q}}^2}{m_{\chi_a^+}^2}\right)
-c_{\tilde{t}}^2\, F^{(3)}_{7,8}
\left(\frac{m_{\tilde{t}_1}^2}{m_{\chi_a^+}^2}\right)
-s_{\tilde{t}}^2\, F^{(3)}_{7,8}
\left(\frac{m_{\tilde{t}_2}^2}{m_{\chi_a^+}^2}\right)
\right]
\right.
\nn[4pt]
&&+\left.
s_{\tilde{t}}c_{\tilde{t}}\,
\frac{{U}_{a2}{V}_{a2}m_t}{2\sin\beta m_{\chi_a^+}}
\left[
F^{(3)}_{7,8}\left(\frac{m_{\tilde{t}_1}^2}{m_{\chi_a^+}^2}\right)
-
F^{(3)}_{7,8}\left(\frac{m_{\tilde{t}_2}^2}{m_{\chi_a^+}^2}\right)
\right]\right\}
,
\label{C78}
\end{eqnarray}
where ${U}$ and  ${V}$ are matrices which diagonalize  
the chargino mass matrix while $s_{\tilde{t}}$ ($c_{\tilde{t}}$) 
denotes the sine (cosine) of the stop mixing angle. 
For more details and the definition of functions 
$F^{(3)}_{7,8}$ see for example \cite{DeGaGi}. 
We will refer to the contribution  in the second line in the 
above formula as a stop-mixing one 
because it vanishes for vanishing stop mixing. 
The sign of this contribution, relative to the SM prediction, 
is the same as ${\rm sgn} \left(\mu A_t\right)$. 
So, it may lead to problems with the $b\to s \gamma$ branching 
ratio in models with negative $\mu$ because $A_t$ very often 
is negative due to the RGE running. One possible way out is to 
have $A_0$ a few times larger than $M_{1/2}$ at the GUT scale 
because in such a case $A_t$ can be positive at the EW scale. 
As a result, stop-mixing contribution would be negative as 
preferred by phenomenology.

Whether BR$(b\to s \gamma)$ can satisfy the experimental constraints 
with light charginos (as preferred by $(g-2)_{\mu}$) and values 
of $A_0$ in a wide range, depends on the sign and the magnitude of 
the contribution in the first line in (\ref{C78}). This contribution 
will be called the gaugino contribution because it involves the 
gaugino components of the charginos ${V}_{a1}$. Using the 
definitions of the chargino mixing matrices ${U}$ and ${V}$ 
(see e.g.\ \cite{HaKa} with an obvious modification for negative $M_2$)
one can find that the sign of this contribution, relative 
to the SM contribution, is given by ${\rm sgn}\left(-\mu M_2\right)$. 
This is very important for the models considered in this paper. 
We have chosen negative $\mu$ (preferred by the Yukawa unification) 
and negative $M_2$ (required by $(g-2)_\mu$ analysis when $\mu<0$). 
As a result, the gaugino part of the chargino-stop contribution 
to BR$(b\to s \gamma)$ (first line in (\ref{C78})) 
has the sign opposite to that of the SM and the charged 
Higgs contributions. This helps to obtain better agreement 
of the $b\to s \gamma$ branching ratio with the experimental results.

The gaugino part of the chargino-stop contribution has the ``correct''
sign preferred by phenomenology. However, it may really help to 
fulfill the experimental bounds only if its magnitude is not 
negligible. Let us check when this is the case. 
The expression in the square bracket in the first line 
of (\ref{C78}) is 
suppressed by the squark GIM mechanism. Moreover, the first line is 
suppressed with respect to the second one by the factor $M_W/m_t$.
This last suppression is not very strong, a factor of about 1/2, 
and may easily be compensated by other sources. Also the squark GIM 
mechanism may be not very efficient when the stop quark masses, 
$m_{\tilde{t}_{1,2}}$, are substantially smaller than the averaged 
squark mass for the first and second families, denoted by $m_{\tilde{q}}$.

There are two more important differences between the two parts 
of eq.\ (\ref{C78}): The second line is proportional to 
$s_{\tilde{t}}c_{\tilde{t}}$ so its value may be suppressed by a smallness 
of the stop mixing. In addition, the ratio of those two contributions 
is proportional to ${V}_{11}/{V}_{12}$. This ratio may be bigger than 
1 if the lighter chargino $\chi_1^+$ is dominated by the gaugino 
component which happens when $M_2^2$ is smaller than $\mu^2$. 
The value of ${V}_{11}/{V}_{12}$ increases with that of 
$\mu^2/M_2^2$ and can be very large. However, one should 
notice that a very large ratio $\mu^2/M_2^2$ suppresses 
both contributions because of the common factor ${U}_{12}$.

Summarizing: the  gaugino part of the 
chargino-stop contribution to  $b\to s \gamma$ 
(first line in (\ref{C78})) may be comparable to or even more 
important than the stop-mixing part (second line) if some of 
the following conditions are met: 
the stop mixing is small; 
the stops are much lighter than the other up-type squarks; 
the lighter chargino is dominated by the gaugino component 
(${V}_{11}$ substantially bigger than ${V}_{12}$).

It turns out that all three above conditions are typically 
easier to satisfy in the presence of a hierarchy $M_{1/2}\ll m_{16}$. 
First of all, for $M_{1/2}\ll m_{16}$ the RGEs for $m_Q$ and $m_U$ are 
dominated by terms proportional to the Yukawa couplings so the mass 
splitting between the stops and other up-type squarks is maximized. 
Secondly, since RGE running usually results in 
$A_t\sim{\mathcal O}(-M_{1/2})$ the hierarchy $M_{1/2}\ll m_{16}$ naturally 
suppresses the stop mixing angle. These arguments are general and are not 
restricted to models with top-bottom-tau Yukawa unification. On the other 
hand, the fact that for $M_{1/2}\ll m_{16}$ it is easier to obtain the 
lighter chargino dominated by the gaugino component, is specifically 
related to the assumption of top-bottom-tau Yukawa unification. 
This follows from the fact that the requirement of not too large 
gluino-sbottom correction to the bottom mass (\ref{mb_gluinocorr}) 
results in the upper bound for $|\mu/M_2|$:
\begin{equation}
\label{muM2bound}
 |\mu/M_2|\lesssim 0.8\ {\rm max}
\left(1,\left(\frac{m_{\tilde b}}{m_{\tilde{g}}}\right)^2\right)
,
\end{equation} 
where we used the approximate relation $m_{\tilde g}\approx 2M_2$ 
resulting from the one-loop RGEs and the assumed pattern of the 
gaugino masses (\ref{gauginoratio}) at the GUT scale. It is clear 
from the above formula that Yukawa unification implies that the 
lighter chargino is dominated by the higgsino component unless the 
gluino is substantially lighter than the heavier sbottom and this 
is possible only if $M_{1/2}\ll m_{16}$.

The above discussed gaugino part of the chargino-stop contribution 
is very important for our analysis. In the models considered 
in this paper (negative $\mu$ and $M_2$) it is negative 
and so helps to obtain acceptable values of BR$(b\to s \gamma)$ 
for lighter SUSY spectrum. This is very desirable because 
the experimental data on $(g-2)_\mu$ strongly favor light 
muon sneutrino and charginos. Our numerical calculations 
show that it is possible to fulfill the experimental bounds 
simultaneously on $(g-2)_\mu$ and  BR$(b\to s \gamma)$. 
The detailed discussion is given in Section \ref{sec_numerical_results}.

\section{Neutralino relic abundance}
\label{sec_LSP}

Experimental data on BR$(b\to s\gamma)$ and $(g-2)_\mu$ result 
in quite strong constraints on SUSY models. Another important 
information which can be used to restrict the SUSY parameter 
space comes from the dark matter (DM) relic
abundance measured by WMAP collaboration
\cite{WMAP7}
\begin{equation}
\Omega_{\rm DM} h^2 = 0.1120 \pm 0.0056\,.
\label{Omega}
\end{equation}
In supersymmetric models with R-parity conservation the lightest
supersymmetric particle (LSP), which is very often the neutralino, 
is stable. In consequence LSP can play the role of DM. 
In the most desirable situation the LSP 
could have the relic abundance within the experimental bounds  
(\ref{Omega}) and so constitute the dominant component of DM. 
Less interesting but still acceptable possibility is the LSP 
relic abundance is below the observed value. 
The neutralino relic abundance depends crucially on the type of 
its dominant component. In models considered in this paper the 
main component of the lightest neutralino is bino. The reason  
is as follows: Due to the pattern of the gaugino masses at the GUT 
scale (\ref{gauginoratio}) and their RG running, the gaugino mass 
ratio at the EW scale is given roughly by 
$|M_1|:|M_2|:|M_3|\approx1:6:12$. This implies that wino cannot be 
the main component of the LSP. Whether bino or higgsino dominates 
LSP  depends on the ratio $|\mu/M_1|$. The requirement of bottom-tau 
Yukawa unification sets the following (conservative) constraint on 
this ratio:
\begin{equation}
 \label{muM1bound}
1.2\lesssim|\frac{\mu}{M_1}|\lesssim 5\,,
\end{equation}
which follows from the requirement that the sbottom-gluino correction 
to the bottom mass (\ref{mb_gluinocorr}) is between 10\% and 20\%  
and from the approximate relation $m_{\tilde{g}}\approx12|M_1|$ 
at the electroweak scale. The above reasoning shows that 
the LSP is mainly bino but in some cases the higgsino component 
may be non-negligible.

Unfortunately, for the bino-like LSP, typical neutralino relic abundance 
is much bigger then (\ref{Omega}). Approximate analytical formulae 
for $\Omega_{\rm DM} h^2$ can be found e.g.\ in \cite{welltempered}. 
One can identify three situations 
when the neutralino relic abundance is not too large: 
First: there are some relatively light superpartners which 
couple strongly enough to the LSP and to some SM particles.
Second: there is a superpartner only slightly heavier than 
the LSP allowing for effective co-annihilation. 
Third: there is a particle approximately two times heavier 
than the LSP coupled strongly enough to the LSP and to some 
SM particles allowing for a resonance enhancement of the 
annihilation cross section. None of these three possibilities 
is very easy to realize. Lower experimental bounds 
on SUSY particles masses make it more and more difficult 
to use the first, simplest, way of decreasing the LSP 
relic abundance. Two other methods, co-annihilation and 
resonance annihilation, require some sort of tuning the 
LSP mass with the mass of some other superpartner. 
So, the condition that the LSP relic abundance is below 
the upper experimental bound results usually in quite 
substantial reduction of the allowed part of the SUSY 
parameter space. 
Our numerical results show that this is the case also 
for the models considered in this paper. Quite big 
parts of the parameter space allowed by all other 
constrains give too large $\Omega_{\rm DM}h^2$. 
Nevertheless, we will show that it is possible to 
fulfill the upper bound on the LSP relic abundance. 
Moreover, each of the three above mentioned mechanism 
may be used in our model. There are regions in which 
the neutralino annihilation cross section is large enough 
due to co-annihilation (mainly with stau), resonance 
annihilation (via $h^0$, $A^0$ or $Z$ boson exchange) or 
presence of light superpartners (e.g.\ sbottom).

\section{Numerical results}
\label{sec_numerical_results}

In order to analyze quantitatively our model we solved numerically the 
2-loop renormalization group equations implementing proper REWSB and 
calculated the sparticle spectrum using SOFTSUSY  \cite{softsusy}  
interfaced with MicrOmegas \cite{Micromega} for calculating the relic 
density of dark matter, as well as, BR$(b\to s \gamma)$, $(g-2)_{\mu}$ 
and BR$(B_s\to\mu^+\mu^-)$. We use the following values of relevant 
experimental inputs: $m_{t}=173.3$ GeV, $m_b(m_b)=4.2$ GeV, 
$\alpha_s(M_Z)=0.1187$.

It is convenient to use the following quantity:
\begin{equation}
 R\equiv\left.\frac{\max\left(h_t,h_b,h_{\tau}\right)}
{\min\left(h_t,h_b,h_{\tau}\right)}\right|_{\rm GUT} 
\end{equation}
to quantify the goodness of Yukawa unification.
In a search for solutions with $R$ close to unity we randomly scanned  
the parameter space defined by:
\begin{equation}
 \begin{tabular}{lll}
$0\leqslant m_{16}\leqslant2000\,{\rm GeV}$ \,, \qquad &
$0\leqslant M_{1/2}\leqslant2000\,{\rm GeV}$ \,, \qquad &
$0.1\leqslant m_{10}/m_{16}\leqslant2$ \,,
\\[4pt]
$0\leqslant D/m_{16}^2\leqslant0.3$ \,, \qquad &
$-3\leqslant A_0/m_{16}\leqslant3$ \,, \qquad &
$40\leqslant \tan\beta\leqslant55$  
 \end{tabular}
\end{equation}
and $\mu<0$. The above ranges for the parameters were chosen for the 
following reasons: First of all, we scanned $m_{16}$ and $M_{1/2}$ only 
below $2$ TeV because we are most interested in solutions which predict 
light enough spectrum which explain the $(g-2)_{\mu}$ anomaly and
could be detected at the LHC. Secondly, we scanned only over positive values 
of $D$ since negative ones seems to be inconsistent with top-bottom-tau 
Yukawa unification. Large values of $\tan\beta$ are necessary to produce 
the observed ratio of top to bottom masses. Ranges for the remaining 
parameters were chosen wide enough not to miss any solutions giving 
Yukawa unification.

For every randomly generated point we demand proper REWSB and the 
neutralino being LSP. We also apply the following experimental constraints:
\begin{align}
 & 12.7\cdot10^{-10}<\delta a_{\mu}^{SUSY}<44.7\cdot10^{-10}\quad (2\sigma)  
\label{exp_g2}\\
 & 2.89\cdot10^{-4}<{\rm BR}(b\to s\gamma)<4.21\cdot10^{-4} \quad (2\sigma) 
\label{exp_bsg}\\
 & {\rm BR}(B_s\to\mu^+\mu^-)<5.8\cdot10^{-8} \\
 &  \Omega_{\rm DM}h^2<0.1288 \quad (3\sigma) 
\label{exp_Omega}\\
 &  m_{h^0}>111.4\,{\rm GeV}
\end{align}
and the mass limits on SUSY particles from LEP and Tevatron. In particular, 
we used the lower bound for the gluino mass $m_{\tilde{g}}>220$ GeV \cite{pdg}.
The impact of the LHC searches for SUSY particles will be discussed later on.

Several comments on the above choice of constraints are in order. 
We imposed only the upper bound on the relic density as a necessary 
consistency condition. We will show later a few benchmark solutions for 
which the WMAP bound on $\Omega_{\rm DM}h^2$ is saturated by the neutralino 
contribution. Due to the fact that the uncertainty in the prediction of 
the lightest Higgs boson mass is about $3$ GeV \cite{Allanach_higgs} we 
used slightly relaxed LEP2 bound \cite{LEPbound}. We use the upper bound on BR$(B_s\to\mu^+\mu^-)$ from Tevatron \cite{BsmumuTeV}. 
 We imposed the $2\sigma$ bound on 
the SUSY contribution to $(g-2)_{\mu}$. So, we demand from the theory 
to really explain the $(g-2)_{\mu}$ anomaly rather than demand only that 
the SUSY contribution to $(g-2)_{\mu}$ is positive (to do no worse than 
the SM) which is often done in the literature. This will allow us to 
understand better the implications of the $(g-2)_{\mu}$ constraint.

\begin{figure}
  \begin{center}
      \resizebox{0.73\textwidth}{!}{\includegraphics{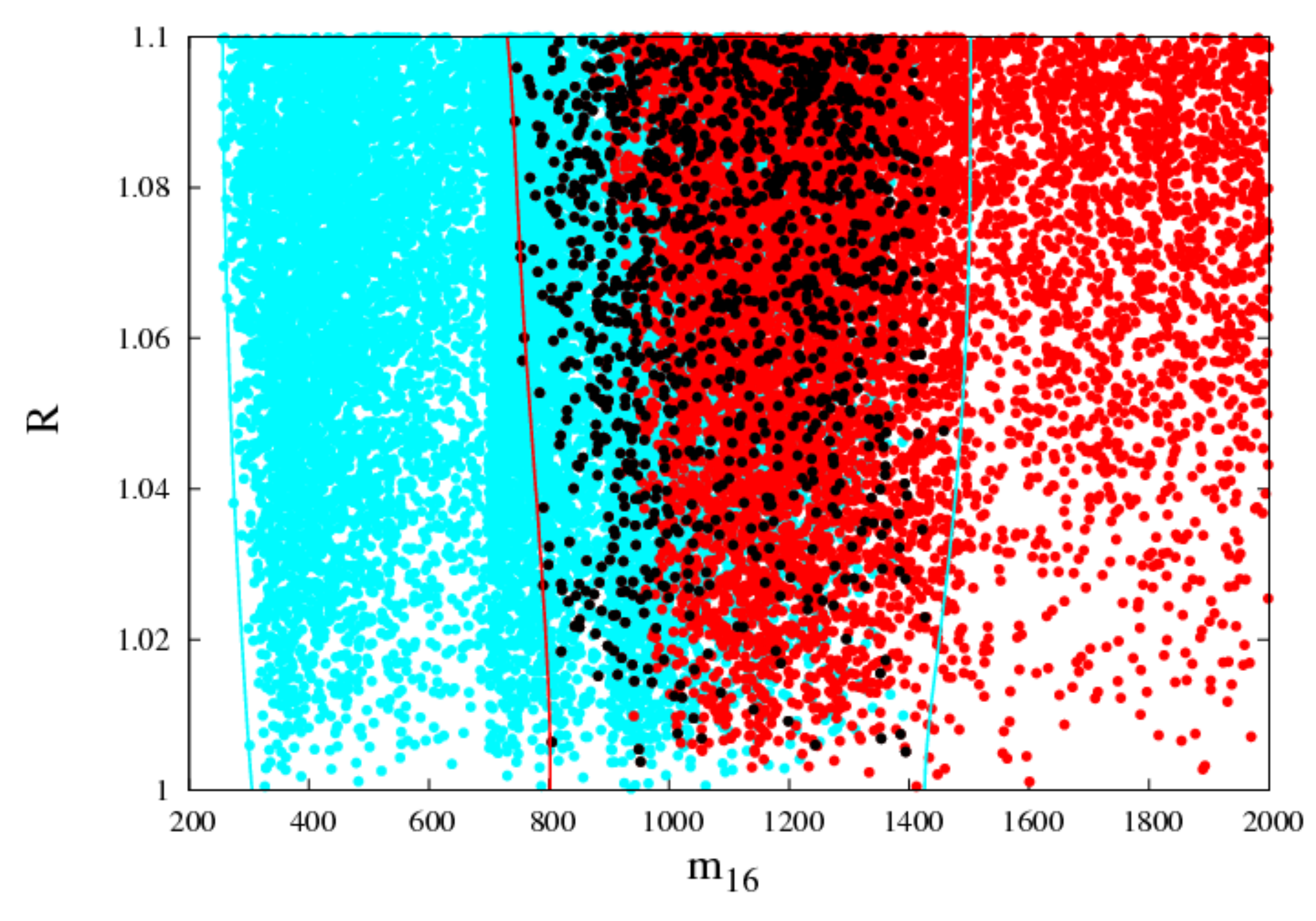}}
    \caption{Plot of $R$ versus $m_{16}$. The regions inside the red (blue) contour are consistent with the constraint from $b\to s \gamma$ ($(g-2)_{\mu}$). In the overlapping region of these contours $b\to s \gamma$ and $(g-2)_{\mu}$ can be satisfied simultaneously. The red (blue) points beside $b\to s \gamma$ ($(g-2)_{\mu}$) satisfy the bound (\ref{exp_Omega}) for $\Omega_{DM} h^2$. The black points satisfy all the constraints including $b\to s \gamma$,  $(g-2)_{\mu}$ and the upper WMAP bound.}
    \label{Rm16}
  \end{center}
\end{figure}

In what follows we present the results of our numerical analysis. 
As we said earlier, for several reasons it is interesting to check  
how constraining for the Yukawa coupling unification are separately 
$(g-2)_{\mu}$ and BR$(b\to s\gamma)$ experimental results, as well as 
the WMAP bound on $\Omega_{\rm DM}h^2$. In Figure \ref{Rm16} we present 
a plot of $R$ vs $m_{16}$. Perfect top-bottom-tau Yukawa unification 
(i.e.\ $R=1$) can be obtained for $m_{16}\gtrsim 300$ GeV. The main 
experimental constraints that restrict possible values of $m_{16}$ 
are $(g-2)_{\mu}$ and BR$(b\to s\gamma)$. For the values of $m_{16}$ 
between about $250$ and $1500$ GeV $(g-2)_{\mu}$ may be within
the $2\sigma$ experimental bound and Yukawa unification may be obtained 
at least at the 10\% level (the region between blue contours in 
Figure \ref{Rm16}). On the other hand, Yukawa-unified solutions 
giving correct values of BR$(b\to s\gamma)$ (at the $2\sigma$ level) 
are found for $m_{16}\gtrsim700$ GeV (the region to the right of the 
red contour in Figure \ref{Rm16}). 
The most interesting fact which follows from our numerical analysis 
is that there exists a rather wide range of values of $m_{16}$, between 
about $700$ and $1500$ GeV, giving a good agreement with the assumption 
of top-bottom-tau Yukawa unification (the overlapping region of the 
red and blue contours in Figure \ref{Rm16}).

\begin{figure}
  \begin{center}
\resizebox{0.73\textwidth}{!}{\includegraphics{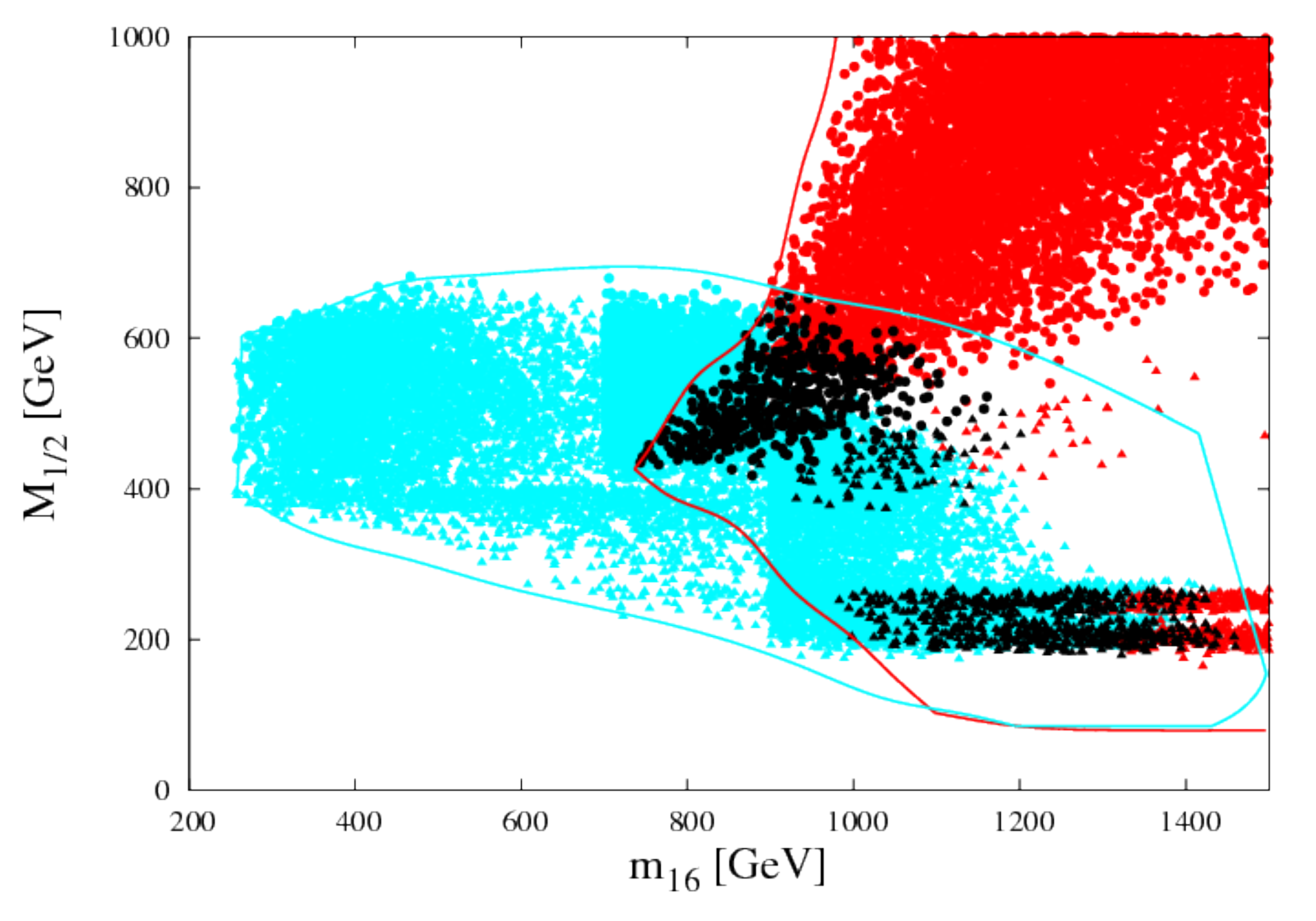}} 
    \caption{Plot of $M_{1/2}$ versus $m_{16}$ assuming $R\leqslant1.1$. The key for the colors of the contours and the points is the same as in Figure \ref{Rm16}. Triangles denote the points with $(m_{\tilde{\tau}}-m_{LSP})/(m_{\tilde{\tau}}+m_{LSP})>0.1$ i.e. solutions which satisfy WMAP bound without stau co-annihilations.}
    \label{m12m16}
  \end{center}
\end{figure}

The WMAP bound on the relic density of neutralinos does not constrain 
the values of $m_{16}$ consistent with top-bottom-tau unification. 
However, this does not mean that the requirement of not too large relic 
abundance of neutralinos does not constrain the parameter space at all. 
In Figure \ref{m12m16} solutions with top-bottom-tau Yukawa unification 
at the level of $10\%$ or better (i.e. $R\leqslant1.1$) are shown in 
the $m_{16}-M_{1/2}$ plane. Without imposing the WMAP bound (\ref{exp_Omega}) 
Yukawa-unified solutions are found inside the blue contour if they 
satisfy the constraint (\ref{exp_g2}) on $(g-2)_{\mu}$ and to the right 
of the red contour if they satisfy the constraint (\ref{exp_bsg}) on 
BR$(b\to s\gamma)$. In the overlapping region of the blue and red contours 
constraints on both $(g-2)_{\mu}$ and BR$(b\to s\gamma)$ can be satisfied 
simultaneously. Even though Yukawa unification with acceptable values 
of $(g-2)_{\mu}$ and BR$(b\to s\gamma)$ can be found for a rather wide range 
of $M_{1/2}$ between about $100$ and $650$ GeV, in almost the entire part of 
the parameter space with $M_{1/2}$ below about $350$ GeV some overabundance of 
neutralinos is predicted. Viable solutions with low $M_{1/2}$ are found 
only in two narrow strips around $M_{1/2}\approx200$ GeV and 
$M_{1/2}\approx250$ GeV which correspond to the mass of the LSP around 
$45$ and $55$ GeV, respectively. Such masses result in very efficient 
resonant annihilation of neutralinos through $Z$ boson or the light CP-even 
Higgs exchange. As seen from Figure \ref{m12m16} these resonances are 
very narrow if the $b\to s\gamma$ constraint (\ref{exp_bsg}) is satisfied. 
This follows from the fact that the couplings of the LSP to $Z$ and $h^0$ 
grow with increasing Higgsino component of the LSP. Therefore, the resonances 
are wider for smaller values of $|\mu/M_1|$. This, however, implies also 
smaller values of $|\mu/M_2|$ which are disfavored by $b\to s \gamma$, 
as discussed in Section \ref{sec_bsg}. The tension between the constraints 
from $\Omega_{\rm DM}h^2$ and $b\to s\gamma$, (\ref{exp_Omega}) and 
(\ref{exp_bsg}), results in a very small range of values of $M_{1/2}$ 
consistent with both observables.

For higher values of $M_{1/2}$ the Yukawa-unified solutions giving 
acceptable values of $(g-2)_{\mu}$ and BR$(b\to s\gamma)$ satisfy the WMAP 
bound usually because of a rather high degree of degeneracy between the 
LSP and the stau which makes the co-annihilation with stau very efficient. 
This occurs for $M_{1/2}$ between about $400$ and $650$ GeV. In order to be 
compatible with the  BR$(b\to s\gamma)$ constraint one needs also 
$A_0/m_{16}\sim{\mathcal O}(2)$ to make the stop-mixing part of the 
chargino contribution to $b\to s\gamma$ negative, or at least strongly 
suppressed (see discussion in Section \ref{sec_bsg}). This is in contrast 
to the Yukawa-unified solutions which satisfy the WMAP bound due to the 
resonant annihilation through $Z$ or $h^0$ for which the stop-mixing part 
of the chargino contribution is suppressed because of a large hierarchy 
$M_{1/2}\ll m_{16}$ so the constraint on BR$(b\to s \gamma)$ can be 
satisfied for any sign of $A_0$.

It can be also seen from Figure \ref{m12m16} that there exist solutions 
consistent with the $(g-2)_{\mu}$ and BR$(b\to s\gamma)$ bounds which 
satisfy the WMAP bound even without a quasi-degeneracy of the LSP and 
stau. They tend to have larger ratio $m_{16}/M_{1/2}$ and 
somewhat smaller values of $M_{1/2}$ than the solutions with the stau
co-annihilation. These solutions can be divided into two subclasses. 
In the first one, the LSP annihilation through $A^0$ is efficient 
because $m_{A^0}$ is relatively light, with masses between about $300$ 
and $400$ GeV, while the LSP mass is about $100$ GeV. It seems far away 
from the center of the $A^0$-resonance but this resonance is very broad 
because the coupling of $A^0$ to $b\bar{b}$ pairs is enhanced by large 
values of $\tan\beta\sim{\mathcal O}(50)$. Notice, however, that such 
a light $A^0$ implies that the contribution to $b\to s\gamma$ from the 
charged Higgs (which is almost degenerate with $A^0$) is very large 
and has to be (at least partially) canceled by a large negative chargino 
contribution. Such cancellation is possible because large positive values 
of $A_0/M_{1/2}\sim{\mathcal O}(5)$ lead to $A_t>0$ at the EW scale. 
Moreover, large values of $A_0/M_{1/2}$ drive stops masses to smaller 
values through RGEs. As a result, the chargino contribution to 
$b\to s\gamma$ is large and negative and may cancel the charged 
Higgs contribution. A large value of $A_0/M_{1/2}\sim{\mathcal O}(5)$ is 
also crucial for the second subclass of solutions consistent with all 
the constraints. Such large values of $A_0/M_{1/2}$ drive the mass of 
the right-handed sbottom (which is always the lightest squark due to 
the negative $D$-term contribution) to smaller values via the RG running 
of $m_D$. We found some Yukawa-unified solutions consistent with 
$(g-2)_{\mu}$, BR$(b\to s\gamma)$ and DM relic abundance constraints 
with a very light sbottom, below about $200$ GeV, for which the neutralinos 
annihilate very efficiently to $b\bar{b}$ pairs through the t-channel 
sbottom exchange. In such a case, there is no need for light $A^0$ nor 
for quasi-degeneracy between the stau and the LSP.

From the previous discussions it should be clear that large positive 
values of $A_0$ help to satisfy the experimental constraints. It is 
interesting to note, however, that there is an upper limit on $A_0/m_{16}$ 
even without taking the experimental constraints into account. 
The reason is as follows: Larger values of the $A$-terms increase 
the value of $\mu^2$ which, in turn, makes the sbottom-gluino contribution 
to the threshold correction to the bottom mass (see eq.\ (\ref{mbsusycorr})) 
more negative. Too large negative SUSY threshold correction to the 
bottom mass worsens bottom-tau Yukawa unification. We found numerically 
that Yukawa unification is impossible for $A_0\gtrsim2.6 m_{16}$.

\begin{figure}
  \begin{center}
\resizebox{0.73\textwidth}{!}{\includegraphics{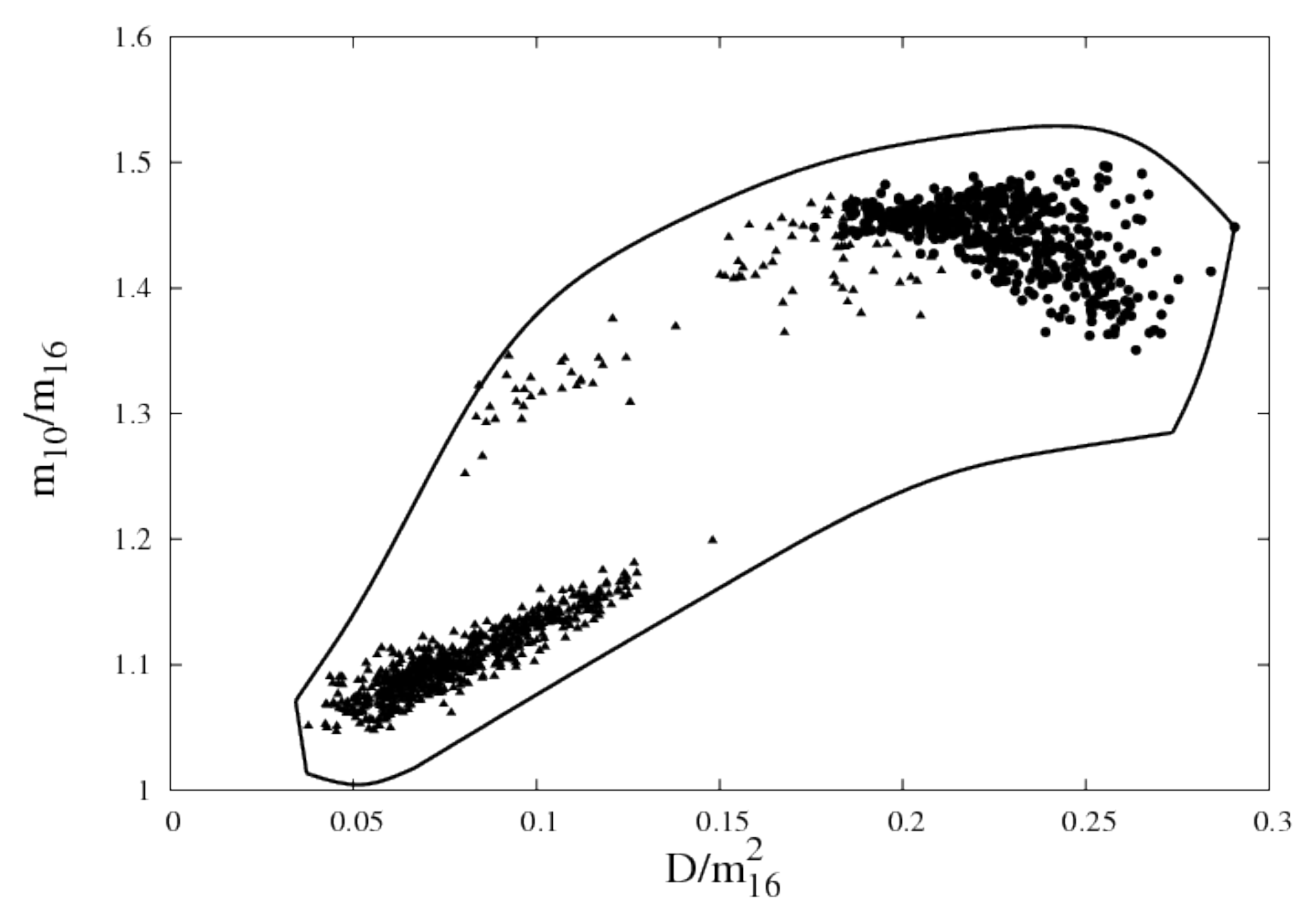}} 
    \caption{Plot of $m_{10}/m_{16}$ vs $D/m_{16}^2$ for solutions with $R\leqslant1.1$ satisfying simultaneously $b\to s \gamma$ and $(g-2)_{\mu}$. The solutions satisfying WMAP bound are denoted by points while without imposing this bound they can be found anywhere inside the contour. Triangles denote the solutions with $(m_{\tilde{\tau}}-m_{LSP})/(m_{\tilde{\tau}}+m_{LSP})>0.1$.}
    \label{Dm10}
  \end{center}
\end{figure}

We should also add that if one requires that only the $(g-2)_{\mu}$ bound  
(\ref{exp_g2}) is satisfied but does not insist on fulfilling the bound  
(\ref{exp_bsg}) on BR$(b\to s\gamma)$ the constraint on the $m_{16}-M_{1/2}$ 
plane from the relic abundance of neutralinos (\ref{exp_Omega}) is much 
weaker. Indeed, it can be seen from Figure \ref{m12m16} that in such a case 
the values of $M_{1/2}$ from about $200$ GeV up to almost $700$ GeV may 
yield a correct relic abundance of neutralinos. The reason is that without 
imposing the $b\to s\gamma$ constraint the pseudoscalar Higgs $A^0$ can 
be very light (but not arbitrary light due to the constraint from 
BR$(B_s\to\mu^+\mu^-)$) so the annihilation through $A^0$ exchange may be 
efficient enough for smaller masses of the LSP (corresponding to smaller 
values of $M_{1/2}$). However, there are no solutions satisfying the 
$(g-2)_{\mu}$ constraint with a correct relic abundance of the neutralinos 
for $M_{1/2}\gtrsim 300$ GeV and large values of $m_{16}\gtrsim 1300$ 
(the lower bound on $m_{16}$ decreases with increasing $M_{1/2}$).

One can infer from Figure \ref{Dm10} that in order to have top-bottom-tau 
Yukawa unification the ratio $D/m_{16}^2$ has to be necessarily positive 
and larger than at least a few hundredths. It is also interesting to note 
that Yukawa unification seems to prefer $m_{10}>m_{16}$ because only 
$m_{10}>m_{16}$ may be consistent with $|\mu|\lesssim M_{1/2}$, as required 
for bottom-tau unification (unless $m_{\tilde{b}_2}>m_{\tilde{g}}$). Let us 
explain this point in more detail. At large $\tan\beta$, the condition of 
proper REWSB implies $\mu^2\approx-\left(m_{H_u}^2+M_Z^2/2\right)$. 
Using this relation and the RGEs one can estimate electroweak scale 
value of $\mu^2$ in terms of the input parameters at the GUT scale:
\begin{equation}
\label{mucoeff_focus}
 \mu^2\approx M_{1/2}^2\left[1.2 +0.65 x^2\left(0.97-\frac{m_{10}^2}{m_{16}^2}
+2.9 \frac{D}{m_{16}^2}+0.15\frac{A_0^2}{m_{16}^2}
-\frac{0.3}{x}\frac{A_0}{m_{16}}\right)\right] ,
\end{equation}
where $x\equiv m_{16}/M_{1/2}$. In order to satisfy the bound (\ref{mubound}) 
the contribution from the gaugino masses to $\mu^2$ has to be (partially) 
canceled by other terms in (\ref{mucoeff_focus}). Yukawa unification 
requires positive $D$-terms so they give an additional positive contribution 
to $\mu^2$. Large values of $|A_0|/M_{1/2}$ also tend to make 
$\mu^2/M_{1/2}^2$ larger. So, partial cancellation in (\ref{mucoeff_focus}) 
may occur only for $m_{10}>m_{16}$. Since $\mu^2$ cannot be negative, 
Yukawa unification consistent with REWSB requires correlations among  
$m_{10}$, $D$ and $A_0$. These correlations are especially strong when 
$M_{1/2}\ll m_{16}$ because in such a case the value of $\mu^2$ is very 
sensitive to the value of the expression in the round bracket in 
eq.\ (\ref{mucoeff_focus}). The correlation between $m_{10}/m_{16}$ 
and $D/m_{16}^2$ is clearly visible in Figure \ref{Dm10}.

\subsection{Predictions for the MSSM spectrum}

A great advantage of the assumption that top, bottom and tau Yukawa 
couplings unify at the GUT scale is its predictivity. When we supplement 
the condition of Yukawa unification by the requirement that 
BR$(b\to s \gamma)$ and $(g-2)_{\mu}$ are consistent with the experimental 
data at the $2\sigma$ level and the relic abundance of neutralinos 
agrees with observations, the predictions for the MSSM spectrum become 
even stronger. This follows mainly from the fact that in such a case 
the values of $M_{1/2}$ and $m_{16}$ are rather tightly constrained. 
The gluino mass is predicted to be between about $500$ and $700$ GeV 
or $900$ GeV and $1.6$ TeV. The gap is the consequence of the fact 
that there are no efficient LSP annihilation channels for values of 
$M_{1/2}$ corresponding to the gluino mass in range between 
about $700$ and $900$ GeV.

The predictions for other sparticles are also rather definite. 
The masses of the heavier stop and sbottom are predicted to be between 
about $800$ GeV and $1.4$ TeV, while the lighter stop has mass typically 
below about $100-200$ GeV. The right-handed sbottom is always lighter 
than other squarks because of the negative $D$-term contribution to its 
mass and typically has the mass between about $300$ and $800$ GeV. 
However, we have found solutions with the right-handed sbottom even as 
light as about $100$ GeV. The squarks of the first two generations are 
typically $300-500$ GeV heavier than the corresponding squarks of the 
third generation because of negligible values of the corresponding 
Yukawa couplings.

\begin{figure}
  \begin{center}
      \resizebox{0.68\textwidth}{!}{\includegraphics{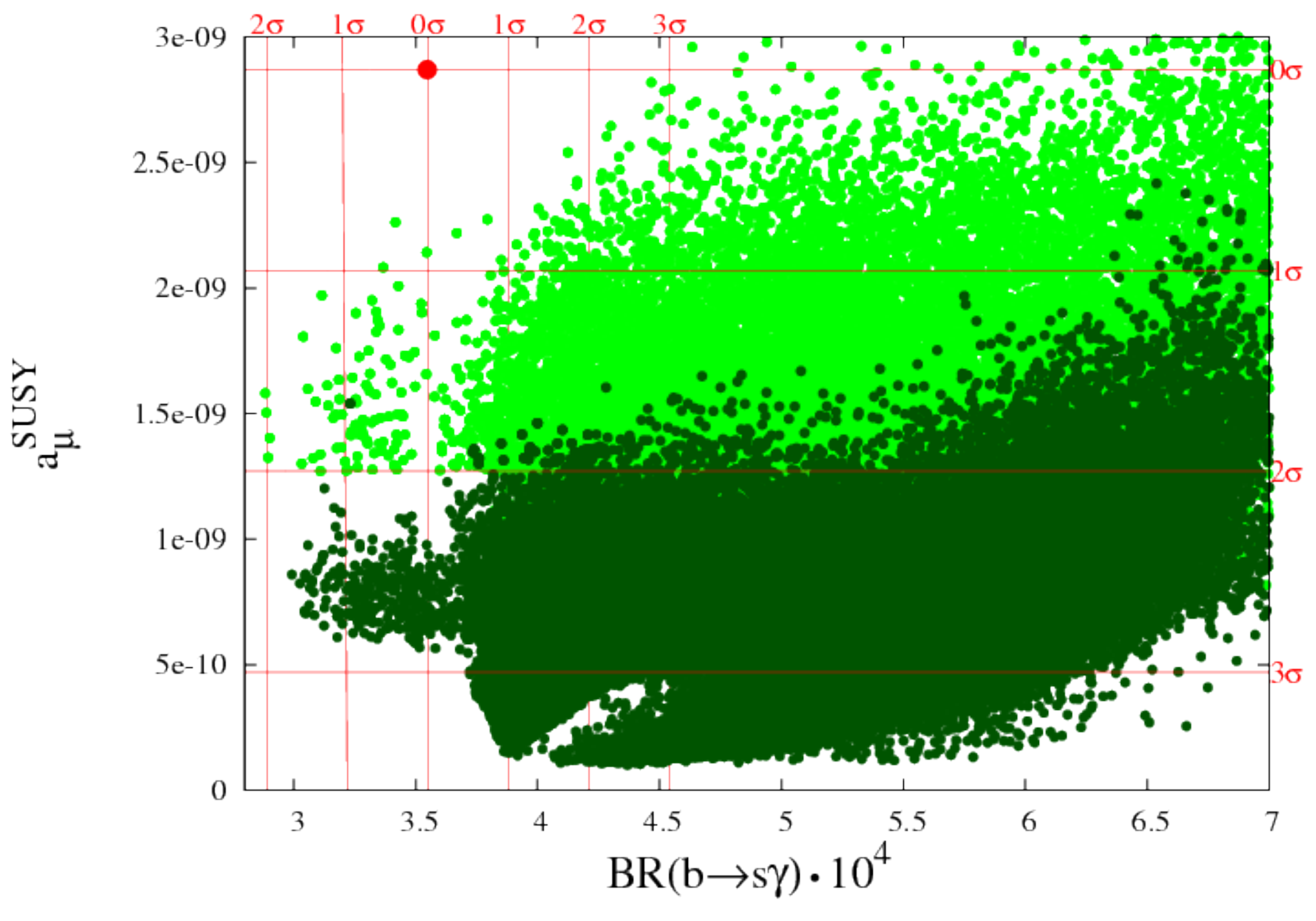}} 
    \caption{Plot of $a_{\mu}^{SUSY}$ versus BR$(b\to s \gamma)$ for the points with $R\leqslant1.1$ and satisfying all the experimental constraints. For light (dark) green points $m_{h^0}>111.4$ ($m_{h^0}>114$) GeV. }
    \label{bsgvsg2}
  \end{center}
\end{figure}

Another prediction of our model is the mass of the lightest Higgs very 
close to the LEP2 bound. This is mainly a consequence of the fact that 
measured value of $(g-2)_{\mu}$ favors smaller values of $m_{16}$ 
which give lighter muon sneutrino but this results also in lighter 
stops making the radiative correction to the Higgs mass smaller.    
Moreover, suppressed stop-mixing required by the BR$(b\to s\gamma)$ 
results in an additional reduction of the Higgs mass. Nevertheless, 
as can be seen in Figure \ref{bsgvsg2}, the experimental bound on 
$(g-2)_{\mu}$ can be satisfied even at the $1\sigma$ level without 
violating the constraint (\ref{exp_bsg}) on $b\to s\gamma$ if 
a conservative bound for the Higgs mass $m_h^0>111.4$ GeV is used 
(we recall that the theoretical uncertainty for the lightest CP-even 
Higgs mass is around $3$ GeV). The maximal value of $(g-2)_{\mu}$ 
which may be obtained in our model drops down rather quickly with 
the Higgs boson mass but for $m_h^0>114$ GeV it can be still 
compatible with the experimental data at $2\sigma$ level.

\begin{figure}
  \begin{center}
\resizebox{0.333\textwidth}{0.33\textheight}{\includegraphics{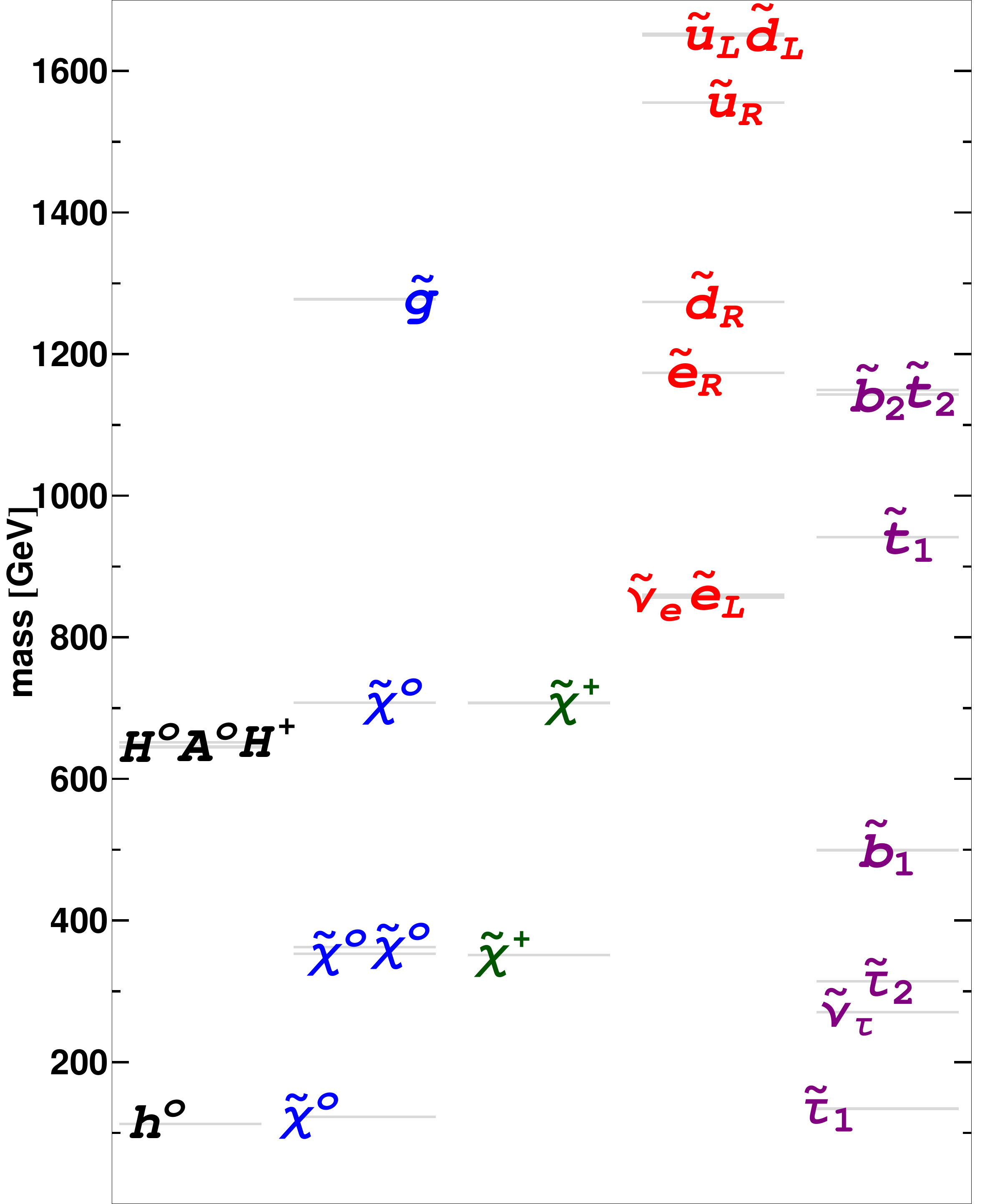}}
\hspace{-11pt}
\resizebox{0.333\textwidth}{0.33\textheight}{\includegraphics{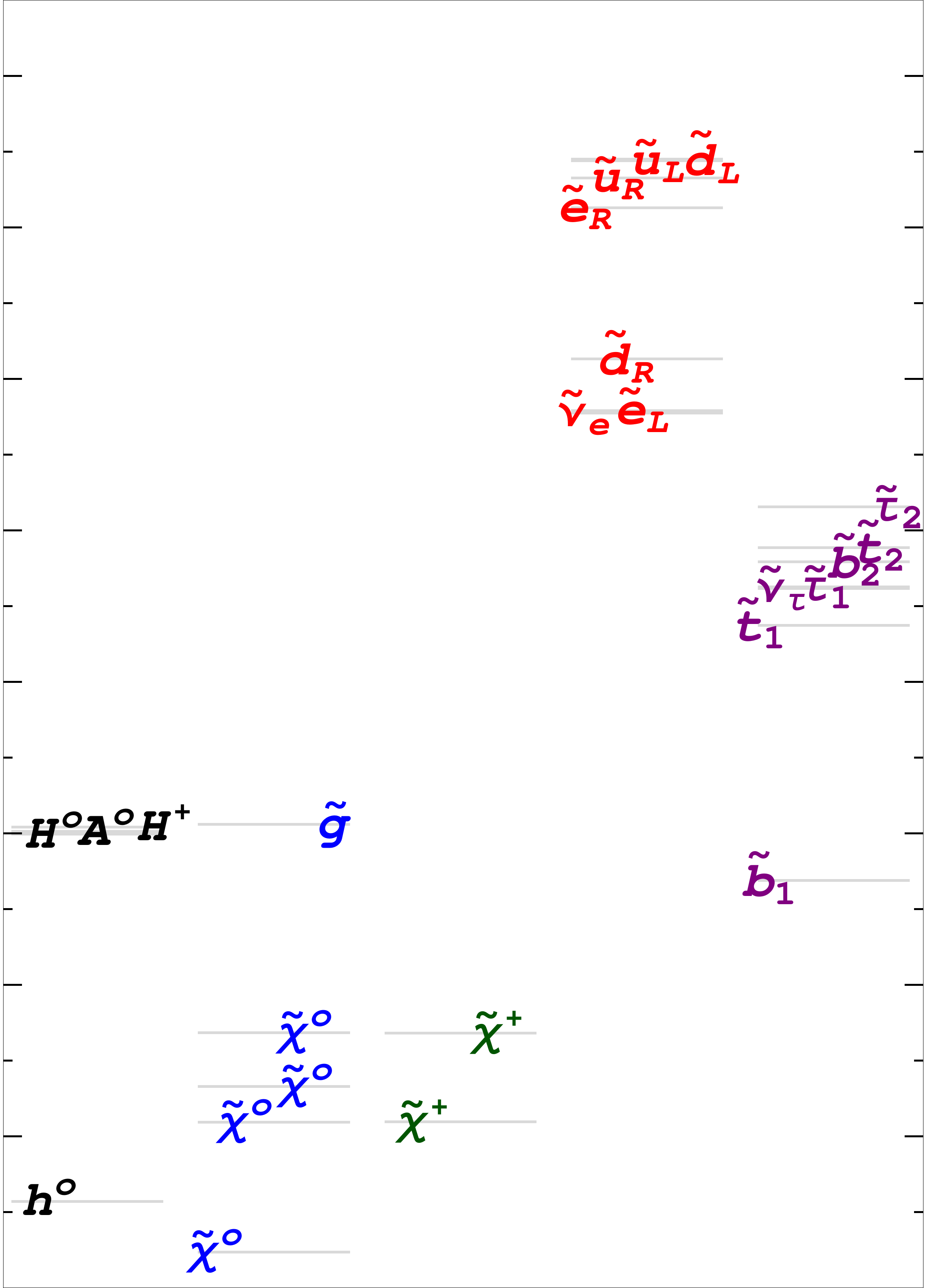}}
\hspace{-9pt}
\resizebox{0.333\textwidth}{0.33\textheight}{\includegraphics{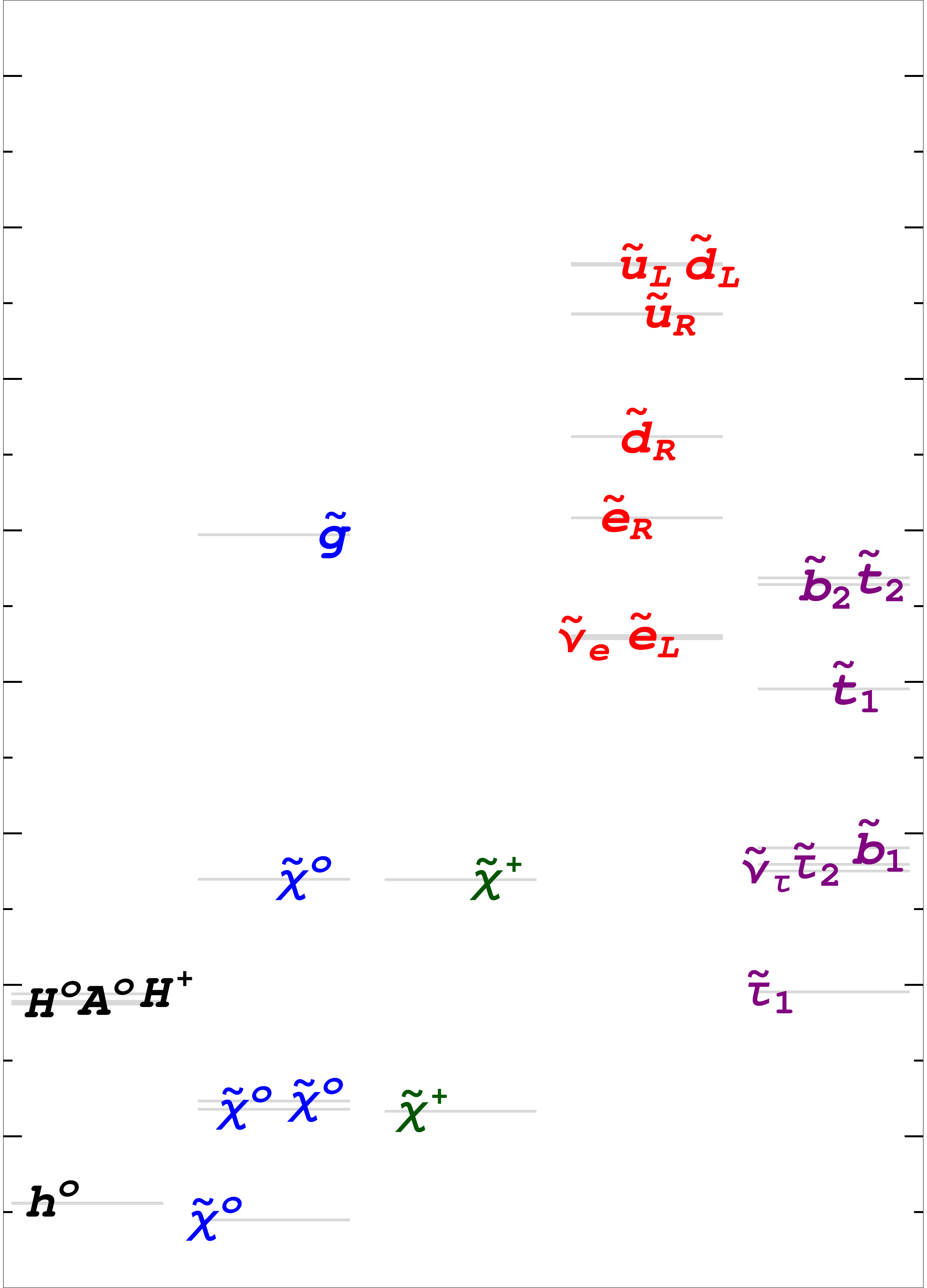}} 
    \caption{Typical spectra in the case of stau co-annihilation (left), resonant annihilation through the Z boson (centre) and the light $A^0$ (right). Sfermions of the second generation (not shown in the Figure) are degenerate with the corresponding sfermions of the first generation. These spectra are obtained for the parameters presented in Table \ref{tabspect} where also values of the relevant observables are given for each case.}
    \label{spectrum}
  \end{center}
\end{figure}

In Figure \ref{spectrum} we present some characteristic spectra 
resulting from the assumption of top-bottom-tau Yukawa unification 
and consistency with all the experimental constraints including 
these on BR$(b\to s \gamma)$ and $(g-2)_{\mu}$.  Different spectra in 
Figure \ref{spectrum} represent classes of solutions which satisfy 
the WMAP bound (\ref{exp_Omega}) due to different main annihilation 
channels of the LSP. Each spectrum has some characteristic features. 
In the case of the stau co-annihilation there is a very large mass 
splitting within each generation of sfermions due to a large value 
of the $D$-term. In the case of the resonant annihilation through the 
Z boson, the gluino is very light. Moreover, since value of $\mu$ can be 
comparable with that of $M_2$ due to the hierarchy $M_{1/2}\ll m_{16}$ 
(see discussion in Section \ref{sec_bsg}) there is a smaller splitting 
in chargino sector and a larger mixing between wino and higgsino. 
This case differs from the others also because it has 
much smaller $|A_0/m_{16}|$ which results in a more compressed spectrum 
of the third generation. Moreover, the pattern of the slepton masses 
in this case reflects the $D$-term splitting at the GUT scale: the 
right-handed stau is heavier than the left-handed slepton doublet. 
On the other hand, in the case of the stau co-annihilation and light 
$A^0$, a large value of $A_0/m_{16}$ makes the left-handed doublet of 
the third generation sleptons heavier than the right-handed stau at the 
EW scale.

In Table \ref{tabspect} the values of the relevant observables 
predicted by these benchmark points are presented. Notice that in each 
case the predicted relic density of neutralinos is within $3\sigma$ 
from the WMAP central value.

\begin{table}
\centering
\begin{tabular}{|l|c|c|c|}
\hline
  & left & centre & right\\ 
\hline
$m_{16}$ & $1054$ & $1352$ & $949.5$ \\ 
$M_{1/2}$ & $542.5$ & $224.2$ & $408.8$\\ 
$m_{10}/m_{16}$ & $1.454$ & $1.125$ & $1.324$\\ 
$D/m_{16}^2$ & $0.1907$ & $0.0921$ & $0.1154$\\
$A_0/m_{16}$ & $2.312$ & $-0.357$ & $1.865$\\
$\tan\beta$ & $45.78$ & $48.47$ & $47.72$\\
\hline
$a_{\mu}^{SUSY}$ & $13.2\times10^{-10}$ & $14.2\times10^{-10}$ & $17.4\times10^{-10}$\\
BR$(b\to s\gamma)$ & $3.83\times10^{-4}$ & $4.11\times10^{-4}$ & $4.19\times10^{-4}$\\
$\Omega_{\rm DM}h^2$ & $0.095$ & $0.111$ & $0.125$\\
\hline
$R$ & $1.007$ & $1.016$ & $1.005$\\
\hline
\end{tabular}
\caption{Input parameters corresponding to SUSY spectra presented in Figure \ref{spectrum} and resulting values of the relevant observables.}
\label{tabspect}
\end{table}

We would like to emphasize that this is the first SO(10) model with 
top-bottom-tau Yukawa unification which predicts a light SUSY spectrum 
with all sparticle masses below $2$ TeV (in some cases even below $1.5$ TeV) 
without violating any experimental constraints. This makes this model 
testable in the near future at the LHC. In fact, since the LHC is 
performing very well it may has already put some constraints on our model. 
The recent ATLAS analysis \cite{gluino_atlas} set the lower bound on the 
gluino mass of about $750$ GeV in a simplified model consisting of gluino, 
degenerate squarks of the first two generations and the massless neutralino. 
However, this bound may be non-applicable to realistic models in which 
the squarks of the third generation must be present. It has been shown 
in \cite{Sakurai} that the limits on the gluino mass from the ATLAS 
experiment are very weak in models with squarks of the third generation 
much lighter than these of the first two generations. In our model the 
EW scale intergenerational mass splitting of the squarks is rather 
significant, especially in the part of the parameter space predicting 
a light gluino between $500$ and $700$ GeV. Therefore, in order to 
set a firm exclusion limits on our model one needs a dedicated 
analysis which we leave for future work.

\section{Conclusions}
\label{sec_conclusions}

We have investigated the SUSY SO(10) GUT model with the negative Higgs 
mixing parameter $\mu$. Non-universal soft SUSY breaking terms have 
been introduced but only those consistent with the SO(10) symmetry. 
Two sources of soft terms generating different masses within 
supermultiplets have been used: the D-terms associated with the breaking 
of SO(10) down to the SM gauge symmetry group in the scalar sector 
and SO(10) non-singlet $F$-terms transforming as the 24-dimensional 
representation of SU(5) $\supset$ SO(10) in the gaugino sector.
No terms explicitly breaking SO(10), like e.g.\ an ad-hoc Higgs 
mass splitting, was allowed. We have shown that top-bottom-tau 
Yukawa unification in such a model is consistent with all 
experimental data. However, the present experimental bounds, 
especially those on BR$(b\to s\gamma)$, $(g-2)_\mu$ and the relic 
dark matter density, are strong and give an important constraints 
on the model parameter space. The main reason is some tension 
between the experimental results and the SM predictions for 
BR$(b\to s\gamma)$ and $(g-2)_\mu$. The simplest way to fulfill 
the bound on $b\to s\gamma$ branching ratio is to have a heavy SUSY 
spectrum. On the other hand, the data on the  muon anomalous magnetic 
moment prefer a light SUSY spectrum. The way to be consistent with the 
constraints on both quantities is to have relatively light 
sparticles but with some special features. This puts constraints 
on the parameter space of the model. The GUT scale value of the 
soft sfermion masses, $m_{16}$, must be in the $800-1400$ GeV range. 
The gluino mass parameter at the GUT scale, $M_{1/2}$, is between about 
100 and 600 GeV with precise bounds depending on $m_{16}$. The values 
of the remaining GUT scale soft parameters, $m_{10}$, $D$ and $A_0$, 
are correlated in order to obtain desirable values of the bottom mass 
threshold correction and of the chargino contribution to $b\to s\gamma$.

The LSP in our model is bino-like. There are several potentially  
important annihilation channels: the co-annihilation 
with stau, the resonance exchange of $Z$ boson or one of the 
Higgses ($h^0$ or $A^0$ if it is light), the t-channel exchange 
of the lighter sbottom. The requirement that the LSP relic abundance 
is below the upper experimental limit restricts further 
the parameter space. There are two allowed ranges of $M_{1/2}$ 
consistent with all experimental constraints. They correspond 
to two ranges of the gluino mass: $500-700$ GeV and $900-1600$ GeV. 
The lower range should be very soon tested in the LHC 
(the lower bound on the gluino mass of about 750 GeV, found 
by ATLAS collaboration in a simplified model, does not apply 
directly to our model).
When the gluino mass is in the upper range, some other particles 
are relatively light with masses below 400 GeV. These may be the 
third generation sleptons (with the lighter stau nearly degenerated 
with the LSP neutralino), the lighter sbottom or the heavier Higgses. 
The LSP neutralino has a mass close to 45 or 55 GeV (when it annihilates 
via the $Z$ or $h^0$ resonance) or between about 80 and 150 GeV 
(when one of other annihilation channels is efficient). 
In addition, at least two other neutralinos and the lighter chargino 
have masses in the $200-400$ GeV range. Even the heaviest sparticles 
are lighter than 2 TeV and quite often lighter than 1.5 TeV. 
Such SUSY spectrum is much lighter than in similar models with 
positive $\mu$. There are good chances that our model can be tested 
by the LHC experiments.

{\bf Note Added:} 
During completion of this work Ref.\ \cite{GoShUn} appeared where 
top-bottom-tau Yukawa unification with $\mu<0$ and non-universal 
gaugino masses was investigated. Even though the same pattern of 
gaugino masses (\ref{gauginoratio}) was considered in \cite{GoShUn} 
there is a major difference between our model and the one studied in 
\cite{GoShUn}. Namely, in \cite{GoShUn} an ad-hoc Higgs mass splitting 
is used which explicitly breaks SO(10) gauge symmetry. In the present paper 
we use the $D$-term splitting of the scalar masses which generically 
arises as a consequence of a spontaneous SO(10) symmetry breakdown.

\section*{Acknowledgments}

This work has received funding from the European Community's Seventh 
Framework Programme under grant agreement PITN-GA-2009-237920 (2009-2013) 
and has been supported by MNiSzW scientific research grant 
N N202 103838 (2010 - 2012). 
MB would like to thank B. C. Allanach, G. Isidori, K. Sakurai 
and especially J. D. Wells for helpful and stimulating discussions.



\begin{thebibliography}{99}

\bibitem{Carena}
  M.~S.~Carena, M.~Olechowski, S.~Pokorski, C.~E.~M.~Wagner,
  Nucl.\ Phys.\  {\bf B426 } (1994)  269-300.
  [hep-ph/9402253].


\bibitem{Olechowski}
  M.~Olechowski, S.~Pokorski,
  Phys.\ Lett.\  {\bf B344 } (1995)  201-210.
  [hep-ph/9407404].


\bibitem{Dterm}
  Y.~Kawamura, H.~Murayama, M.~Yamaguchi,
  Phys.\ Rev.\  {\bf D51 } (1995)  1337-1352.
  [hep-ph/9406245].

\bibitem{Baernegmu}
 H.~Baer, M.~A.~Diaz, J.~Ferrandis, X.~Tata,
  Phys.\ Rev.\  {\bf D61 } (2000)  111701.
  [hep-ph/9907211];
  H.~Baer, M.~Brhlik, M.~A.~Diaz, J.~Ferrandis, P.~Mercadante, P.~Quintana, X.~Tata,
  Phys.\ Rev.\  {\bf D63 } (2000)  015007.
  [hep-ph/0005027].


\bibitem{Martin}
  S.~P.~Martin,
  Phys.\ Rev.\  {\bf D79 } (2009)  095019.
  [arXiv:0903.3568 [hep-ph]].

\bibitem{Blazek}
  T.~Blazek, R.~Dermisek, S.~Raby,
  Phys.\ Rev.\  {\bf D65 } (2002)  115004.
  [hep-ph/0201081].


\bibitem{Auto}
  D.~Auto, H.~Baer, C.~Balazs, A.~Belyaev, J.~Ferrandis, X.~Tata,
  JHEP {\bf 0306 } (2003)  023.
  [hep-ph/0302155].


\bibitem{BaerDM}
  H.~Baer, S.~Kraml, S.~Sekmen, H.~Summy,
  JHEP {\bf 0803 } (2008)  056.
  [arXiv:0801.1831 [hep-ph]].


\bibitem{Baerjustso}
  H.~Baer, S.~Kraml, S.~Sekmen,
  JHEP {\bf 0909 } (2009)  005.
  [arXiv:0908.0134 [hsep-ph]].


\bibitem{ChNath}
  U.~Chattopadhyay and P.~Nath,
  Phys.\ Rev.\  D {\bf 65} (2002) 075009
  [arXiv:hep-ph/0110341].

\bibitem{ChCoNath} 
  U.~Chattopadhyay, A.~Corsetti, P.~Nath,
  Phys.\ Rev.\  {\bf D66 } (2002)  035003.
  [hep-ph/0201001].

\bibitem{wronggaugino}
  N.~Chamoun, C.~S.~Huang, C.~Liu and X.~H.~Wu,
  Nucl.\ Phys.\  B {\bf 624} (2002) 81
  [arXiv:hep-ph/0110332].



\bibitem{GoKhRaSh}
  I.~Gogoladze, R.~Khalid, S.~Raza, Q.~Shafi,
  JHEP {\bf 1012 } (2010)  055. [arXiv:1008.2765 [hep-ph]].

\bibitem{Hall}
  L.~J.~Hall, R.~Rattazzi, U.~Sarid,
  Phys.\ Rev.\  {\bf D50 } (1994)  7048-7065.
  [hep-ph/9306309].


\bibitem{Pierce}
  D.~M.~Pierce, J.~A.~Bagger, K.~T.~Matchev, R.~-j.~Zhang,
  Nucl.\ Phys.\  {\bf B491 } (1997)  3-67.
  [hep-ph/9606211].


\bibitem{Wells_yuk}
  K.~Tobe, J.~D.~Wells,
  Nucl.\ Phys.\  {\bf B663 } (2003)  123-140.
  [hep-ph/0301015].


\bibitem{Murayama}
  H.~Murayama, M.~Olechowski, S.~Pokorski,
  Phys.\ Lett.\  {\bf B371 } (1996)  57-64.
  [hep-ph/9510327].


\bibitem{Davier_g2SM}
  M.~Davier, A.~Hoecker, B.~Malaescu, Z.~Zhang,
  Eur.\ Phys.\ J.\  {\bf C71 } (2011)  1515.
  [arXiv:1010.4180 [hep-ph]].


\bibitem{BNL}
  G.~W.~Bennett {\it et al.} [ Muon G-2 Collaboration ],
  Phys.\ Rev.\  {\bf D73 } (2006)  072003.
  [hep-ex/0602035].


\bibitem{Misiak}
  M.~Misiak, H.~M.~Asatrian, K.~Bieri, M.~Czakon, A.~Czarnecki, T.~Ewerth, A.~Ferroglia, P.~Gambino {\it et al.},
  Phys.\ Rev.\ Lett.\  {\bf 98 } (2007)  022002.
  [hep-ph/0609232].


\bibitem{HFAG}
  D.~Asner {\it et al.}  [Heavy Flavor Averaging Group],
  arXiv:1010.1589 [hep-ex].


\bibitem{Chankowski}
  P.~H.~Chankowski, J.~R.~Ellis, M.~Olechowski, S.~Pokorski,
  Nucl.\ Phys.\  {\bf B544 } (1999)  39-63.
  [hep-ph/9808275].

\bibitem{Okumura}
  K.~-i.~Okumura, L.~Roszkowski,
  Phys.\ Rev.\ Lett.\  {\bf 92 } (2004)  161801.
  [hep-ph/0208101].

\bibitem{Moroi}
T.~Moroi,
  Phys.\ Rev.\  D {\bf 53} (1996) 6565
  [Erratum-ibid.\  D {\bf 56} (1997) 4424]
  [arXiv:hep-ph/9512396].



\bibitem{Stockinger}
  D.~Stockinger,
  J.\ Phys.\ G {\bf G34 } (2007)  R45-R92.
  [hep-ph/0609168].

\bibitem{DeGaGi}
 G.~Degrassi, P.~Gambino and G.~F.~Giudice,
  JHEP {\bf 0012} (2000) 009
  [arXiv:hep-ph/0009337].


\bibitem{HaKa}
H.~E.~Haber and G.~L.~Kane,
  Phys.\ Rept.\  {\bf 117} (1985) 75.


\bibitem{WMAP7}
 E.~Komatsu {\it et al.}  [WMAP Collaboration],
  Astrophys.\ J.\ Suppl.\  {\bf 192} (2011) 18
  [arXiv:1001.4538 [astro-ph.CO]].


\bibitem{welltempered}
  N.~Arkani-Hamed, A.~Delgado, G.~F.~Giudice,
  Nucl.\ Phys.\  {\bf B741 } (2006)  108-130.
  [hep-ph/0601041].

\bibitem{softsusy}
  B.~C.~Allanach,
  Comput.\ Phys.\ Commun.\  {\bf 143 } (2002)  305-331.
  [hep-ph/0104145].

\bibitem{Micromega}
  G.~Belanger, F.~Boudjema, A.~Pukhov, A.~Semenov,
  Comput.\ Phys.\ Commun.\  {\bf 176 } (2007)  367-382.
  [hep-ph/0607059].

\bibitem{pdg}
  K.~Nakamura {\it et al.} [ Particle Data Group Collaboration ],
  J.\ Phys.\ G {\bf G37 } (2010)  075021.


\bibitem{Allanach_higgs}
  B.~C.~Allanach, A.~Djouadi, J.~L.~Kneur, W.~Porod and P.~Slavich,
  JHEP {\bf 0409} (2004) 044
  [arXiv:hep-ph/0406166].

\bibitem{LEPbound}
  S.~Schael {\it et al.}  [ALEPH Collaboration and DELPHI Collaboration and
                  L3 Collaboration and ],
  Eur.\ Phys.\ J.\  C {\bf 47} (2006) 547
  [arXiv:hep-ex/0602042].

\bibitem{BsmumuTeV}
  T.~Aaltonen {\it et al.} [ CDF Collaboration ],
  Phys.\ Rev.\ Lett.\  {\bf 100 } (2008)  101802.
  [arXiv:0712.1708 [hep-ex]].

\bibitem{gluino_atlas}
ATLAS Collaboration, ATLAS-CONF-2011-086

\bibitem{Sakurai}
  K.~Sakurai, K.~Takayama,
  [arXiv:1106.3794 [hep-ph]].

\bibitem{GoShUn}
I.~Gogoladze, Q.~Shafi and C.~S.~Un,
  arXiv:1107.1228 [hep-ph].




\end{thebibliography}
\end{document}